\documentclass[twocolumn,showpacs,preprintnumbers,amsmath,amssymb]{revtex4}
\usepackage{amsfonts}
\usepackage{amssymb}
\usepackage{psfig}
\newcommand{\beq}{\begin{equation}}
\newcommand{\eeq}{\end{equation}}
\newcommand{\beqn}{\begin{eqnarray}}
\newcommand{\eeqn}{\end{eqnarray}}
\newcommand{\bea}[1]{\beq\begin{array}{#1}}
\newcommand{\eea}{\end{array}\eeq}

\newcommand{\ket}[1]{|\,#1\,\rangle}
\newcommand{\bra}[1]{\langle\,#1\,|}
\newcommand{\braket}[2]{\langle\,#1\,|\,#2\,\rangle}

\newcommand{\diff}{\partial}

\newcommand{\cL}{{\cal L}}

\newcommand{\HP}[1]{\mathrm{HP}^#1}
\newcommand{\qq}{\langle qq \rangle}
\begin{document}
\preprint{ITEP-LAT/2005-16}

\title{Lattice Gauge Fields Topology Uncovered
       by Quaternionic $\sigma$-model Embedding}

\author{F.V.~Gubarev}
 \email{gubarev@itep.ru}
\author{S.M.~Morozov}
 \email{smoroz@itep.ru}

\affiliation{Institute of Theoretical and  Experimental Physics,
              B.Cheremushkinskaya 25, Moscow, 117218, Russia}

\begin{abstract}
We investigate SU(2) gauge fields topology using new approach,
which exploits the well known connection between SU(2) gauge theory and
quaternionic projective $\sigma$-models and allows to formulate the topological
charge density entirely in terms of $\sigma$-model fields.
The method is studied in details and for thermalized vacuum configurations
is shown to be compatible with overlap-based definition.
We confirm that the topological charge is distributed in localized
four dimensional regions which, however, are not compatible with instantons.
Topological density bulk distribution is investigated at different lattice
spacings and is shown to possess some universal properties.
\end{abstract}

\pacs{11.15.-q, 11.15.Ha, 12.38.Aw, 12.38.Gc}

\maketitle
\section{Introduction}

It is hardly possible to overestimate the significance of the topological fluctuations
in the vacuum of Yang-Mills theories. The discovery of instantons~\cite{Polyakov},
merons~\cite{Callan} and other topologically non-trivial solutions
was the clue to understand the major non-perturbative phenomena in QCD
and to construct rather successful phenomenological low-energy models
(see, e.g., Ref.~\cite{Schafer:1996wv} and references therein). 
At the same time the validity of these models was in fact questioned from very
beginning~\cite{Witten:1978bc} and in particular because in the strong coupling regime
they become unreliable with no control on the degree of approximation made.

The unique approach to investigate the issue would be the numerical lattice simulations.
However, until mid 90's all lattice methods  aimed to investigate the topological aspects
of the gauge theories were plagued by essentially the same decease since they could
not be applied reliably to 'hot' vacuum configurations. Here we mean the
field-theoretical~\cite{DiVecchia:1981qi}
and geometrical~\cite{Luscher,Stone} definitions of the topological charge on the lattice which are known
to be an art rather then the unambiguous first-principle investigation tools (see, e.g., Ref.~\cite{Pugh}).
The situation was ameliorated when the lattice chiral fermions were constructed~\cite{Neuberger:1997fp}
and a nice  alternative topological charge definition was provided~\cite{overlap} via
Atiyah-Singer index theorem.

Since then a lot of data had been accumulated and we're not in the position
to review the current status of the problem.
Instead we note that strictly speaking the usual field-theoretical topological charge 
and the one constructed from fermionic modes are equivalent in the continuum limit
and for smooth fields only. In the context of regularized quantum field theory
they are {\it a priori} distinct and are sensitive quite differently to the details
of regularization. It is believed that the continuum limit of both definitions
is the same, but at finite cutoff the overlap-based construction is much more advantageous
although it is not ideal either, in particular, because of its complexity.

The purpose of this paper is to develop an alternative approach to the topology
of $D=4$ pure SU(2) lattice gauge theory, which ideologically is very close to
what had been said above. Namely, we propose to consider the definition
of the topological charge which is known to be equivalent to the standard one in
the continuum limit. However, at finite cutoff it becomes an independent construction
and for this reason is worth to be investigated.
Our approach could be best illustrated in the context of
$D=2$ $\mathrm{CP}^1$ (or $O(3)$) $\sigma$-model where the topological charge
\beq
Q_{\mathrm{CP}^1} = \frac{1}{8\pi} \int d^2 x \, \varepsilon^{\mu\nu} \,
\vec{n} \cdot [\diff_\mu \vec{n} \times \diff_\nu \vec{n}]
\eeq
could be identically rewritten in terms of U(1) gauge potentials $A_\mu = -i z^* \diff_\mu z$
\beq
Q_{\mathrm{CP}^1} = \frac{1}{4\pi} \int d^2 x \, \varepsilon^{\mu\nu} \, F_{\mu\nu}\,,
\eeq
where $F_{\mu\nu}$ is the usual Abelian field-strength and complex $z$ is related to $\vec{n}$
by standard stereographic projection. The idea is to follow the above
equations in the opposite order and formulate the topological charge of $D=2$ U(1)
gauge fields as the corresponding one in $\mathrm{CP}^n$ $\sigma$-model, where we have
indicated that in fact the $\sigma$-model rank $n$ is a free parameter.
The above reasoning generalizes almost trivially to SU(2) case. The difference
is that we have to consider the field of real quaternions $\mathrm{H}$ instead of the
complex numbers and the corresponding $D=4$ $\sigma$-model target space
is the quaternionic projective space $\HP{n}$.
Note that the deep connection between $\HP{n}$ $\sigma$-models and SU(2) Yang-Mills
theory is in fact well known (see, e.g.,
Refs.~\cite{Gursey:1979tu,HPN-reviews,Dubrovin} and references therein).
In particular, the geometry of gauge fields could best be analyzed
in the $\HP{n}$ context. In fact, all known instantonic
solutions of Yang-Mills theory could be induced
from the topological configurations of suitable $\HP{n}$ $\sigma$-model.

Thus it is natural to expect that the construction of 'nearest' to the given gauge background
$\HP{n}$ fields captures accurately the gauge fields topology
leaving aside their non-topological properties. 
Clearly, the notion of 'nearest' is crucial and it is described in details below.
Here we note that it inevitably introduces a sort of ambiguity in our approach the significance
of which is hardly possible to analyze {\it a priori} for quantum vacuum configurations.
Although for classical gauge fields the corresponding $\HP{n}$ $\sigma$-model is known to be unique,
the mathematical rigour is lost in case of 'hot' vacuum fields and the corresponding
systematic errors are to be investigated separately, presumably using numerical methods.
This concerns, in particular, the topological charge defined via $\HP{n}$ $\sigma$-model embedding
and therefore its properties near the continuum limit are to be carefully studied.
These issues are considered only preliminary in present publication
and therefore the values of physical quantities quoted below are to be taken with care.
Note, however, that
our results do indicate that the ambiguity of $\HP{n}$ $\sigma$-model construction
for equilibrium vacuum configurations is likely to be irrelevant while 
other systematic errors could easily be reduced algorithmically.

The paper is organized as follows. In section~\ref{HPN-intro} we give basic
theoretical background, introduce our notations and describe the proposed
approach in the continuum terms. Section~\ref{HPN-lattice} is devoted
to the detailed description and investigation of our method on the lattice.
In particular, we consider the dependence of the topological charge and its density
definitions upon the $\HP{n}$ $\sigma$-models rank and other parameters involved.
Then the algorithm is tested on semiclassical fields and compared with overlap-based
topological charge on 'hot' SU(2) configurations, where we found a quite remarkable agreement.

Section~\ref{HPN-topology} is devoted exclusively to the investigation of SU(2) gauge fields
topology as it is seen by our method and a few closely related issues.
We show in section~\ref{HPN-dynamics} that the dynamics of embedded
$\HP{1}$ $\sigma$-model closely reflects the dynamics of the original gauge fields,
in particular, the mass gap of the $\sigma$-model is
given by SU(2) string tension. Next in section~\ref{global-Q} the topological susceptibility is considered and
we find that it scales rather precisely being in agreement with the existing
literature. Then in section~\ref{top-density} we investigate in details the local
structure of the topological charge
at the particular value $0.1193(9) \mathrm{~fm}$ of the lattice spacing.
We confirm that the topological charge bulk distribution has rather
peculiar lumpy structure discovered in previous studies~\cite{Gattringer,DeGrand,Hip,Horvath:2002gk}
which, however, has nothing to do with instantons.
In particular, the lumps are distributed according to $N(V_\cL) \sim V_\cL^{-3/2}$,
where $V_\cL$ is the lump 4-volume,
and are dominated by UV small lumps. Nevertheless, the topological susceptibility
turns out to be lumps saturated.
It is amusing that setting the parameters involved in the lumps definition to unphysical
(as we believe) values we are able to qualitatively reproduce the recently
discovered~\cite{Horvath-structures}
global topological structures (section~\ref{limit}). Namely, the lumps organize themselves into
a few (typically two) percolating structures per configuration plus the divergent amount of UV small
lumps consisting mainly of just one lattice point. However, for reasons which we discuss in details
it is unclear for us what is the physical significance of this result.
Then in section~\ref{qq-corr} the topological density correlation function
is considered and we argue that it should not
be necessary negative within our approach. Instead it reflects the lumpy structure mentioned above
and the corresponding correlation length is determined by the characteristic lump size, which
turns out to be of order half the lattice spacing.
Finally, in section~\ref{scaling} we confront the data obtained at two different $\beta$ values
and argue that the parameters
involved in the lumps definition are likely to be given in physical units. Moreover, the lumps
volume distribution $N(V_\cL) \sim V_\cL^{-3/2}$ seems to be spacing independent.
As far as the characteristic lump size is concerned, we are not confident in
its scaling properties, more data is needed to quantify the issue. 

\section{$\HP{n}$ $\sigma$-models and their Embedding into SU(2) Yang-Mills theory}
\label{HPN-intro}

The purpose of this section is to outline in the continuum terms the approach we proposed
to investigate SU(2) gauge fields topology. In particular, we briefly remind the
construction of quaternionic projective spaces $\HP{n}$ and corresponding
four dimensional $\sigma$-models. The material of this section is by no
means new, for review and more details see, e.g., Refs.~\cite{Gursey:1979tu,HPN-reviews,Dubrovin}.

\subsection{Quaternionic Projective Spaces and $\HP{n}$ $\sigma$-models}
\label{HPN-theory}

$\HP{n}$ is an example of quaternionic Grassmann manifold and is
a compact symmetric space of quaternionic dimensionality $n$ on which
the group $Sp(n+1)$ acts transitively. It can be viewed as the factor space
\beq
\HP{n} ~=~ \frac{Sp(n+1)}{ Sp(n) \times Sp(1)}
\eeq
and therefore for arbitrary $n$ we have
\beq
\label{homotopy}
\pi_4(\HP{n}) = \pi_3(Sp(1)) = \pi_3(SU(2)) = Z\,.
\eeq
Another useful representation of $\HP{n}$ is
\beq
\label{HPN-sphere}
\HP{n} = S^{4n+3} / S^3\,,
\eeq
hence $\HP{n}$ is the fibering of $S^{4n+3}$ over quaternionic lines passing through
origin. In the particular case $n=1$, which will be the most important below, we have
\beq
\HP{1} ~=~ S^7 / S^3
\eeq
(second Hopf fibering). The simplest explicit parametrization of 
$\HP{n}$ is as follows. Consider normalized quaternionic vectors
\beq
\label{q-vector}
\ket{q} = [q_0, ... , q_n]^T\,, \qquad
\braket{q}{q} = \sum\limits_{i=0}^n \bar{q}_i q_i = 1 \in \mathrm{H}\,,
\eeq
where $[...]^T$ denotes transposition, $q_i = q_i^{\alpha} e_\alpha$, $\alpha = 0,...,3$
are real quaternions (homogeneous coordinates) with $e_\alpha$ being the quaternionic units
\beq
e_0 = 1\,,\quad
e_i e_j = -\delta_{ij} + \varepsilon_{ijk} e_k\,, \,\,\,\,i,j,k = 1,2,3
\eeq
and quaternion conjugation is defined by
\beq
\bar{q}_i = q_i^{\alpha} \bar{e}_\alpha\,, \qquad
\bar{e}_0 = e_0\,,\,\,\,\, \bar{e}_i = -e_i\,,\,\,\,\, i=1,2,3\,.
\eeq
Therefore the states $\ket{q}$ describe the $(4n+3)$-dimensional sphere, $\ket{q} \in S^{4n+3}$.
According to Eq.~(\ref{HPN-sphere}) the $\HP{n}$ space is the set of equivalence classes of
vectors $\ket{q}$ with respect to the right multiplication by unit quaternions (right
action of SU(2) gauge group)
\beq
\label{gauge-inv}
\ket{q} \sim \ket{q} \, \upsilon\,,\qquad |\upsilon|^2 \equiv \bar{\upsilon} \upsilon = 1\,,
\quad \upsilon\in\mathrm{H}\,.
\eeq
As usual it is convenient to introduce quaternionic $(n+1) \times (n+1)$ projection matrices
\beq
\label{projector}
P = \ket{q}\bra{q} \quad \left( P_{ik} = q_i \bar{q}_k \right)\,, \qquad
P^2 = P = P^\dagger\,,
\eeq
which are invariant under (\ref{gauge-inv}) and therefore parametrize $\HP{n}$.
Alternatively, one could consider the matrix
\beq
N = 2 P - 1\,,\qquad N = N^\dagger\,,\,\,\,\, N^2 = 1\,,
\eeq
which generalizes to $\HP{n}$ case the familiar stereographic projection of $S^4=\HP{1}$. Indeed,
for $n=1$ the two-component vector (\ref{q-vector}) is uniquely defined by
inhomogeneous coordinate $\omega = q_1 q^{-1}_0 \in \mathrm{H}$ and 
in terms of unit five dimensional vector $n^A$
\beq
\label{projection}
n^\alpha = \frac{2 \omega^\alpha}{ 1 + |\omega|^2}\,, \,\,\,\,\alpha=0,...,3\,,
\qquad
n^4 = \frac{1 - |\omega|^2}{1 + |\omega|^2}
\eeq
we have $N = \sum_{A=0}^4\gamma^A n^A$, where $\gamma^A$ are the five Euclidean Dirac matrices
$\{\gamma^\mu,\gamma^5\}$.

As far as the $\HP{n}$ $\sigma$-models are concerned it is clear from Eq.~(\ref{homotopy})
that the model is non-trivial provided that 
the base space is taken to be the 4-sphere $S^4$. The index
of the mapping $S^4 \to \HP{n}$ is given in terms of the projectors (\ref{projector}) by
\beq
\label{Q-P}
Q = \frac{1}{4\pi^2} \int d^4 x \,\, \varepsilon^{\mu\nu\lambda\rho} \,\,
\mathrm{Sc} \, \mathrm{tr}\left( P \diff_\mu P \diff_\nu P \diff_\lambda P\diff_\rho P\right)\,,
\eeq
where $\mathrm{tr}$ means the trace over quaternionic indices and scalar part is defined by
$\mathrm{Sc}\,q = (q + \bar{q})/2$. The action of the model is given also in terms of
projectors (\ref{projector}),
but its explicit form will not be needed in what follows. Note that Eq.~(\ref{Q-P})
simplifies greatly in $n=1$ case. In terms of five dimensional unit vector (\ref{projection})
the topological charge is
\beqn
\label{Q-n}
Q = \frac{1}{(8\pi)^2} \int d^4 x \,\, \varepsilon^{\mu\nu\lambda\rho}
    \, \varepsilon_{ABCDE} \cdot ~~~~~~~~~~~~~~~~~~~ \\
~~~~~~~~~~~~~~~~~~~ \cdot n^A \diff_\mu n^B \diff_\nu n^C \diff_\lambda n^D\diff_\rho n^E \nonumber
\eeqn
and geometrically is the sum of the oriented infinitesimal volumes in the image
of the mapping $n^A(x) : S^4 \to S^4$.

An alternative way to describe the $\HP{n}$ $\sigma$-models is to consider auxiliary
SU(2) gauge fields $A_\mu$ transforming as usual
$A_\mu \to \bar{\upsilon} A_\mu \upsilon - \bar{\upsilon}\diff_\mu\upsilon$
under (\ref{gauge-inv}). The covariant derivative
\beq
D\ket{q} = \diff \ket{q} + \ket{q} A
\eeq
transforms homogeneously $D\ket{q} \to D\ket{q} \upsilon$ and both 
the topological charge and the action
could be written equally in terms of $D\ket{q}$. The potentials $A_\mu$
are not dynamical degrees of freedom and could be eliminated in favor of $\ket{q}$
\beq
\label{gauge-fields}
A_\mu = - \bra{q} \diff_\mu \ket{q} = - \sum\limits_{i=0}^n \bar{q}_i \diff_\mu q_i\,.
\eeq
However, it is crucial that the topological charge (\ref{Q-P}), (\ref{Q-n}) is expressible
solely in terms of $A_\mu$
\beqn
\label{Q-A}
Q = \frac{1}{32\pi^2} \int d^4 x \,\, \varepsilon^{\mu\nu\lambda\rho}\,\,
\mathrm{Tr} \, F_{\mu\nu} F_{\lambda\rho}\,, \\
F_{\mu\nu} = \diff_\mu A_\nu - \diff_\nu A_\mu + [A_\mu, A_\nu]\,, \nonumber
\eeqn
being essentially equivalent to the familiar topological charge of the gauge fields
(\ref{gauge-fields}). A particular illustration is provided
by hedge-hog configuration of $n^A$ fields of $\HP{1}$ $\sigma$-model. One can show
that indeed Eq.~(\ref{gauge-fields}) leads in this case to the canonical instanton
solution of SU(2) Yang-Mills theory.

\subsection{Relation to SU(2) Yang-Mills Theory}
\label{HPN-method}

The difference between the $\HP{n}$ $\sigma$-models and the usual
SU(2) Yang-Mills theory lies in the fact the gauge potentials (\ref{gauge-fields})
are not independent but composite fields constructed in terms of elementary scalars.
However, this difference becomes less significant due to the theorem by
Narasimhan and Ramanan~\cite{theorem} (see also~\cite{Dubrovin}). Namely, it was shown that any sourceless
SU(2) gauge fields considered on a compact manifold can be induced from the canonical
connection of suitable Stiefel bundle (it goes without saying that the theorem applies
to smooth fields only). In particular, that means that any classical SU(2)
gauge potentials on $S^4$ could be realized it terms of $\HP{n}$ $\sigma$-fields
as in Eq.~(\ref{gauge-fields}) for sufficiently large but finite $n$. 
It should be noted that in fact the representation (\ref{gauge-fields}) is well known and was used,
in particular, in Ref.~\cite{Atiyah:1978ri} to construct the general instanton solutions
in Yang-Mills theory. Moreover, one could formulate~\cite{Dubois-Violette:1978it}
the gauge theories  entirely in terms of projectors (\ref{projector}).

Notice also a particular logic underlying the considerations above. Namely, one starts
from the $\HP{n}$ $\sigma$-fields the topology of which could be analyzed either directly
with Eq.~(\ref{Q-P}) or indirectly with Eq.~(\ref{Q-A}) when the gauge fields (\ref{gauge-fields})
are introduced. In the continuum limit once the differentiability is assumed
these two approaches are essentially the same. However, it is crucial that in
quantum field theory regularized with finite UV cutoff~\footnote{
Here we mean mainly the lattice regularization with lattice spacing $a$.
} these two approaches need not be identical anymore. 
Moreover at finite cutoff various methods are expected to be sensitive quite differently to the
details of regularization and, in particular, to UV-scale fluctuations and lattice artifacts.
The well known example of this kind is provided by the topological
charge (\ref{Q-A}) in Yang-Mills theory for which the field-theoretical~\cite{DiVecchia:1981qi},
geometrical~\cite{Luscher,Stone} and Atiyah-Singer index theorem based~\cite{Seiler,overlap}
constructions are quite distinct at finite $a$ (for review and further references see,
e.g., Ref.~\cite{topo-review}).

Since our ultimate goal is to investigate the topology of Yang-Mills fields at finite
UV cutoff, we could try to reverse the above logic and attempt to use
Eqs.~(\ref{Q-P}), (\ref{Q-n}) in the context of gauge theories.
Therefore generically our approach looks as follows.

{\it i)} For given SU(2) gauge potentials and for given rank $n$
one finds the corresponding closest $\HP{n}$ $\sigma$-model fields
for which the approximation (\ref{gauge-fields}) is as best as possible.
The quality of approximation {\it a priori} depends upon the rank
and deserve a separate investigation.

{\it ii)} Then the topology of the gauge fields could be analyzed
in terms of $\HP{n}$ $\sigma$-model. In particular, the topological charge density
is given by oriented four-dimensional volume of infinitesimal tetrahedron in the
image of the map $P: S^4\to \HP{n}$ provided by projectors (\ref{projector}).

It is clear that the actual implementation of the above program is challenging,
there are a wealth of technical issues to be discussed below. However, right now
let us mention the following.

{\it a)} Comparing the number of degrees of freedom one concludes that the
approximation (\ref{gauge-fields}) could not be exact for small rank $n$. 
At the same time it is well known that the classical instanton solution
in Yang-Mills theory is exactly reproduced by $\HP{n}$ $\sigma$-model already
for $n=1$. This means, in particular, that for gauge fields representing
classical instanton plus perturbative fluctuation the above construction
for $n=1$ is likely to reproduce correctly the gauge fields topology while being almost blind
to the perturbative noise. Naively, one could expect that the approximation (\ref{gauge-fields})
is to be more stringent as rank is increased. Then it follows that the rank of the $\HP{n}$ $\sigma$-model
could play the role of ultraviolet filter: the larger the rank $n$ the better is
the approximation (\ref{gauge-fields}) and therefore the above approach becomes
more sensitive to the ultraviolet fluctuations. 

{\it b)} In the continuum limit the topological charge density is given by the oriented
volume of infinitesimal 4-dimensional tetrahedron embedded into $\HP{n}$ space. 
However, at any small but finite lattice spacing the tetrahedron need not be
infinitesimal and its vertices could be far away from each other (in the standard $\HP{n}$
metric). The commonly accepted prescription in this case (see, e.g., Refs.~\cite{Luscher,Stone,Berg,Gerrit})
is to connect  vertices pairwise by shortest geodesics and calculate the 4-volume
of corresponding 'geodesic' tetrahedron. Unfortunately, it seems that this procedure
is impossible to implements practically for $n>1$. Indeed, to the best of our knowledge
even in the simplest case $n=1$ the analytical expression for the volume of
4-dimensional spherical tetrahedron embedded into $S^4$ space does not exist.
Moreover, for $n>1$ we don't see any practical way to calculate the volume even
numerically. However, for $n=1$ the simple geometry of the space $\HP{1} = S^4$
admits a reasonable numerical solution (see below for details).
Therefore in the investigation of the gauge fields topology we are forced to
consider the $\HP{1}$ $\sigma$-model only. However, in view of the item {\it a)}
above this should not be considered as fatal restriction of our method
(see also the next section).

\section{Lattice Implementation}
\label{HPN-lattice}

In this section we describe in details the implementation of the above approach
on the lattice. Section~\ref{HPN-embedding} is devoted 
to the problem of $\sigma$-model embedding for given SU(2) lattice gauge fields background.
In particular, we discuss the dependence of approximation (\ref{gauge-fields}) upon
the rank $n$ and investigate the Gribov copies issue. In section~\ref{HPN-top-density}
we describe the topological charge density calculation method which uses solely
the $\HP{1}$ $\sigma$-model fields. 

\subsection{$\HP{n}$ $\sigma$-models Embedding on the Lattice}
\label{HPN-embedding}

The construction of $\HP{n}$ $\sigma$-model starts from assigning quaternionic
vectors $\ket{q_x}$, Eq.~(\ref{q-vector}),
to each lattice site and then the actual problem  is how Eq.~(\ref{gauge-fields})
translates to the lattice. The l.h.s. of Eq.~(\ref{gauge-fields}) should evidently
be given by link matrices $U_{x,\mu}$. As far as the r.h.s. is concerned 
its lattice counterpart is in principle ambiguous. In fact, the correct lattice
replacement of Eq.~(\ref{gauge-fields}) could be found by analogy with corresponding case
of $\mathrm{CP}^n$  $\sigma$-models~\cite{Berg} or deduced from considerations
of quaternionic quantum mechanics~\cite{Adler}
\beq
\label{approx1}
U_{x,\mu} ~=~ \frac{\braket{q_x}{q_{x+\mu}}}{ |\braket{q_x}{q_{x+\mu}}|}\,.
\eeq
Since generically it is not possible to satisfy Eq.~(\ref{approx1}) identically
the best what we can do is to minimize the functional
\beqn
\label{approx}
 & F = \frac{1}{V} \sum\limits_x \delta F_x\,, & \\
 & \delta F_x = \frac{1}{4} \sum\limits_\mu ( 1 - 
\mathrm{Sc} [ U^\dagger_{x,\mu} \frac{\braket{q_x}{q_{x+\mu}}}{|\braket{q_x}{q_{x+\mu}}|}] )
& \nonumber
\eeqn
over all $\{\ket{q_x}\}$ configurations (here $V$ denotes the lattice volume).
Note that the minimization tasks of this kind are expected to be plagued by Gribov copies
problem to be discussed below.
We implemented the usual local overrelaxed minimization of the functional (\ref{approx}).
The stopping criterion is that the difference of $F$ values between two consecutive sweeps should
be smaller then $10^{-6}$.

The first question to be addressed is the quality of the approximation (\ref{approx1}),
the natural measure of which is given by $\langle\delta F_x \rangle$. We measured
$\langle \delta F_x \rangle$ on about 20 thermalized statistically
independent SU(2) gauge configurations on $12^4$ lattice with $\beta=2.35$
(the rank of the $\sigma$-model was $n = 1$).
The averaged value turns out to be $\langle \delta F_x \rangle \approx 0.12$ (see also below)
which in fact indicates rather good approximation quality.
What is more important is that $\langle \delta F_x \rangle$ appears to be slightly
decreasing function of $\beta$. For instance, on $16^4$ lattice at $\beta=2.40$
we found $\langle \delta F_x \rangle \approx 0.10$.

\begin{figure}[t]
\centerline{\psfig{file=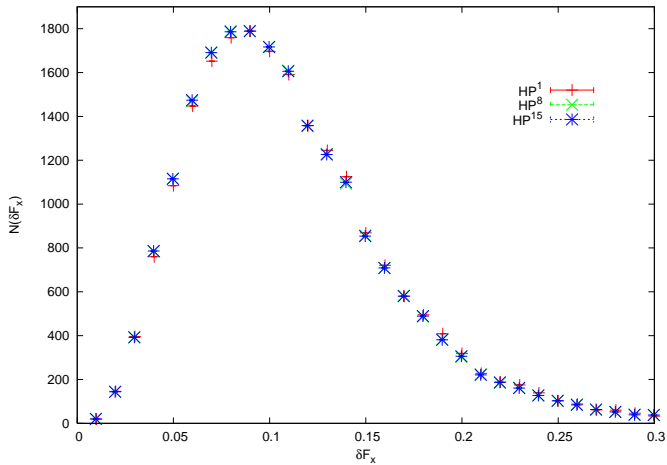,width=0.5\textwidth,silent=,angle=-90,clip=}}
\caption{Typical distribution of $\delta F_x$, Eq.~(\ref{approx}),
on thermalized $12^4$, $\beta=2.35$ lattices for various $\sigma$-model ranks.}
\label{fig:approx_rank_histo}
\end{figure}

Next we have to investigate how the quality of approximation depends upon the $\sigma$-model rank $n$.
We calculated the distribution of $\delta F_x$ for $n=1,8,15$ on several fixed thermalized
SU(2) configurations ($12^4$ lattice at $\beta=2.35$).
The corresponding typical histograms are presented on Fig.~\ref{fig:approx_rank_histo} and clearly show
that the approximation quality does not depend at all upon the rank.
Thus the naive expectation that rank could serve as an ultraviolet filter turns out to be wrong.
Instead we found that the approximation quality is 10-20\%
for 'hot' SU(2) gauge fields irrespectively of the concrete $\HP{n}$ $\sigma$-model considered.
The plausible explanation of this might be as follows: from the continuum considerations we know
that the equality (\ref{gauge-fields}) is valid for smooth (e.g., classical) SU(2) fields only.
It is commonly believed that the configurations at $\beta=2.35$ are not smooth.
For this reason the approximation (\ref{approx1}) could not be exact for any rank
and its quality is mainly given by the gauge fields roughness and indeed rank independent.
Note that if this reasoning is valid then
the approximation quality (\ref{approx}) provides a quantitative measure
of gauge fields smoothness. The conclusion is that it is unnecessary
to consider $\HP{n}$ $\sigma$-models with $n > 1$ and from now on we confine ourselves
to the case $n=1$.

\begin{figure}[t]
\centerline{\psfig{file=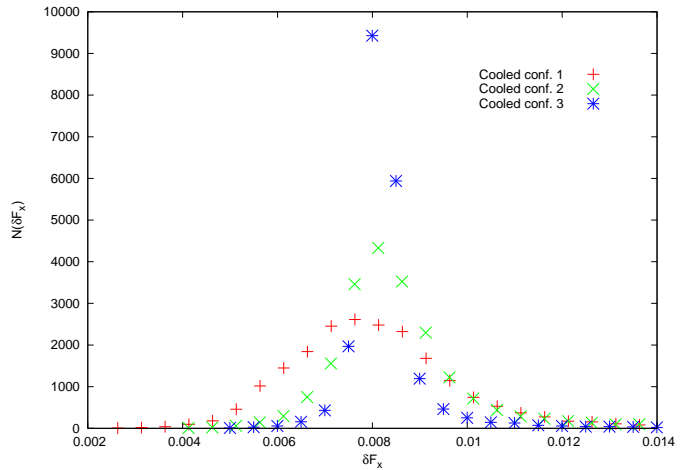,width=0.5\textwidth,silent=,angle=-90,clip=}}
\caption{Distribution of $\delta F_x$, Eq.~(\ref{approx}),
for $\HP{1}$ $\sigma$-model on three highly cooled $12^4$ configurations
containing respectively three instantons, one anti-instanton and no instantons at all.}
\label{fig:approx_ins_histo}
\end{figure}

\begin{figure}[t]
\centerline{\psfig{file=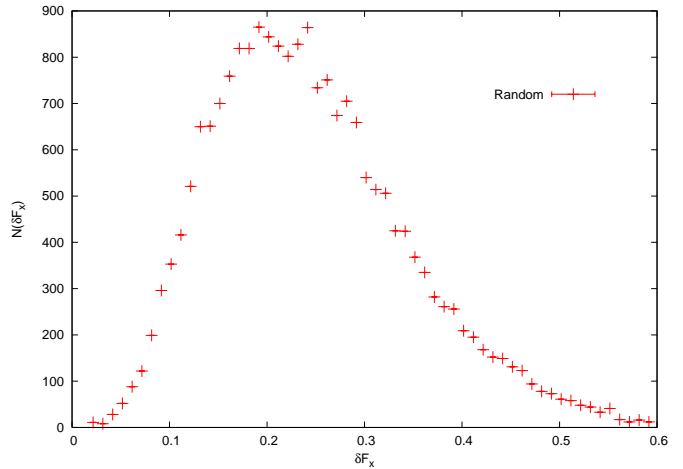,width=0.5\textwidth,silent=,angle=-90,clip=}}
\caption{$\delta F_x$ distribution on $12^4$ lattice for random input gauge fields.}
\label{fig:approx_random_histo}
\end{figure}

In view of the above discussion it is instructive to consider semiclassical and completely random
input gauge fields.
Using the cooling procedure (see, e.g., \cite{GarciaPerez:1998ru}  and references therein)
we generated three highly cooled $12^4$ configurations containing respectively
three instantons, one anti-instanton and no instantons at all. The corresponding distribution
of $\delta F_x$ is shown on Fig.~\ref{fig:approx_ins_histo}. It is worth to mention that
the approximation quality (\ref{approx1}) turns out be excellent, the gauge fields
are reproduced with accuracy $\lesssim$ 1\%
in the whole lattice volume. Note that for non-trivial classical background the distribution
of $\delta F_x$ is slightly broader then that for trivial case still reproducing the input
gauge fields almost exactly. As far as the random gauge fields are concerned,
the results obtained in this case (Fig.~\ref{fig:approx_random_histo}) support the above argumentation.
Namely, the distribution of $\delta F_x$ values is much broader being significantly large
even at $\delta F_x \approx 0.5$. The corresponding quality of approximation could be estimated as
$\langle\delta F_x\rangle \approx 0.3$
and indeed is much worse then that in previous cases.

\begin{figure}[t]
\centerline{\psfig{file=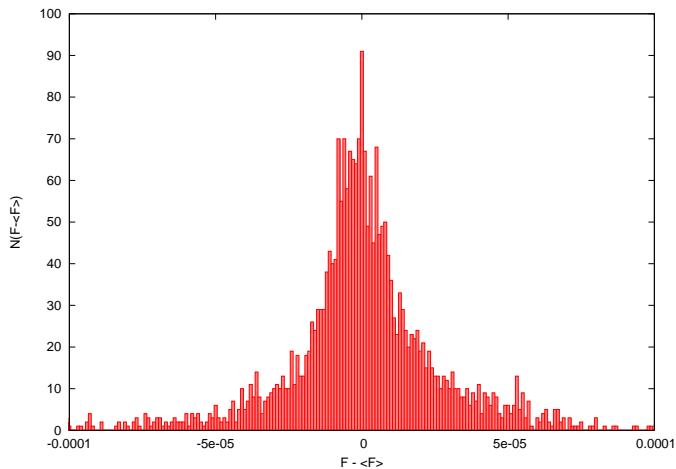,width=0.5\textwidth,silent=,angle=-90,clip=}}
\caption{Distribution of the minimal values of the functional (\ref{approx})
(with mean value subtracted) calculated on $12^4$ $\beta=2.35$ thermalized lattices
and averaged over 20 configurations, see text.}
\label{fig:approx_Gribov}
\end{figure}

The last problem to be considered in this section is the problem of Gribov copies
usually associated with minimization of functionals like (\ref{approx}).
The problem lies in the fact that there might be many local
minima of (\ref{approx}) which are relatively far apart in the configuration space
of $\ket{q_x}$ fields.  To investigate this issue we performed the following calculation. 
For a particular thermalized gauge configuration ($12^4$ lattice at $\beta=2.35$)
the minimization of (\ref{approx}) was repeated 50 times starting every time with
some random $\ket{q_x}$. From the deviations of found minimal values
from their mean the distribution histogram was constructed and
averaged over 20 input gauge configurations. In order to quantitatively address the Gribov copies
problem one should compare the width of $F$ values distribution obtained this way and
the precision of the functional (\ref{approx}) calculation which in our case could be estimated
as $\approx 10^{-6}$.
The distribution of the minimal values of $F$ is shown on Fig.~\ref{fig:approx_Gribov}
and its width turns out to be of order $\approx 10^{-5}$. The conclusion is that the Gribov
copies problem is likely to be inessential in finding the $\HP{1}$ $\sigma$-model for given
SU(2) gauge background. However, for safety reasons we always tried 10 random initial
configurations $\ket{q_x}$ in our minimization procedure and then selected the best minimum
found.

\subsection{Topological Charge Density}
\label{HPN-top-density}

As we already mentioned the topological charge density in terms of the $\HP{1}$
$\sigma$-model fields is given by the oriented 4-volume of spherical tetrahedron $T$ embedded
into $S^4$. The necessity to consider 4-dimensional simplices could be deduced already
from the continuum expression (\ref{Q-n}) which depends upon
unit vectors $n^A_{(i)}$ at five infinitesimally close neighboring points
marked by index $i=0,...,4$. On the lattice
the vertices of the tetrahedron $T = \{n^A_{(i)}\}\in S^4$ become finitely separated
and should be connected pairwise by shortest geodesics thus forming a spherical tetrahedron
embedded into $S^4$.

To the best of our knowledge the volume of spherical four dimensional tetrahedron is unknown
analytically contrary to the case of three dimensions (see, e.g., Refs.~\cite{3tetrahedron}
and references therein). Therefore, the only way to proceed is to invent some reasonable numerical
algorithm to evaluate the 4-volume in question. 
Note that this is so because we are specifically interested in the topological charge
density. As far as only the global topological charge is of concern, one could proceed
in usual way utilizing the fact that the topological density is almost the total derivative
(see, e.g., Ref.~\cite{Gerrit}). Then the evaluation of the global charge reduces
to the calculation of the volumes of three-dimensional spherical simplices
which could be done analytically.

The most straightforward way to estimate the 4-volume of spherical simplex
is to use the Monte Carlo technique, the essence of which is to pick up a random point in
$S^4$ and to decide whether it is inside or outside the spherical tetrahedron.
Due to the simplicity of $S^4$ geometry the last question reduces to the relatively
fast calculation $5\times 5$ determinants analogously to the case of $S^3$ sphere
considered in~\cite{Stone}.
However, proceeding this way we definitely loose the integer valuedness of the
topological charge since volume of every simplex (topological charge density)
is calculated with finite accuracy. Evidently the accuracy should be small enough
to recover the topological charge and simultaneously large enough to be practically
reasonable. In our test runs (see below) we found that the optimal relative accuracy
of the spherical tetrahedron volume estimation should be smaller then 2\%.

Now we can summarize our algorithm of topological charge density calculation.

1. On input there are five unit 5-dimensional vectors $n^A_{(i)}$, $i=0,...,4$ sitting at five
vertices of simplex ${\cal T}$ of the physical space triangulation. 
The vectors $n^A_{(i)}$ are constructed via
Eq.~(\ref{projection}) and provide the mapping of ${\cal T}$ into  the spherical
tetrahedron $T \in S^4$, the oriented volume $V(T)$ of which is to be calculated.
Generically the set $n^A_{(i)}$ is not degenerated $\mathrm{det}_{iA}\,[n^A_{(i)}] \ne 0$.
Note that for $|\mathrm{det}_{iA}\,[n^A_{(i)}]| < 10^{-10}$ we equate $V(T)$ to zero.

2. In order to speed up the Monte Carlo integration we calculate the normalized mean 
vector $N \propto \sum_i n_{(i)}$ and find the 5-dimensional cone $C$ around $N$
containing all the $n_{(i)}$. Since $V(T) < V( C \cap S^4)$ the tetrahedron volume
is equated to zero if $V( C \cap S^4) < 10^{-8}$.

3. The Monte Carlo estimation of $V(T)$ begins with picking up $10^3$ random points
uniformly distributed in $C \cap S^4$. If none of them falls into the interior of
$T$ the tetrahedron volume is equated to zero. Otherwise we continue to generate
uniformly in $C \cap S^4$ random points until  the volume $V(T)$ is known with accuracy 2\%.
Finally, the topological charge density in the simplex ${\cal T}$ of physical space
triangulation is given by
\beq
\label{q-lat}
q({\cal T}) = \frac{3}{8\pi^2} \, \mathrm{sign}\left(\mathrm{det}_{iA}\,[n^A_{(i)}]\right) \cdot V(T)\,,
\eeq
where the normalization factor $8\pi^2/3$ is the volume of $S^4$.
The corresponding topological charge is
\beq
\label{Q-lat}
Q_{float} = \sum\limits_{\cal T} q({\cal T})\,,
\eeq
where we have explicitly indicated that it is not integer valued.
We remark that in order exploit this procedure on usual hypercubical lattices
the hypercubes should be sliced into simplices. We did this according to the
prescription of Ref.~\cite{Kronfeld:1986ts}.

\begin{figure}[t]
\centerline{\psfig{file=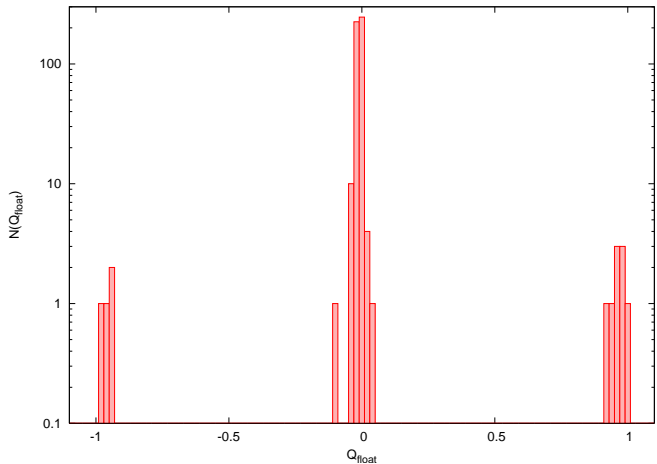,width=0.5\textwidth,silent=,angle=-90,clip=}}
\caption{Distribution of the topological charge $Q_{float}$, Eq.~(\ref{Q-lat}),
for single 5-dimensional simplex with random $\HP{1}$ $\sigma$-model fields at its vertices.}
\label{fig:top_simplex_histo}
\end{figure}

\begin{figure}[t]
\centerline{\psfig{file=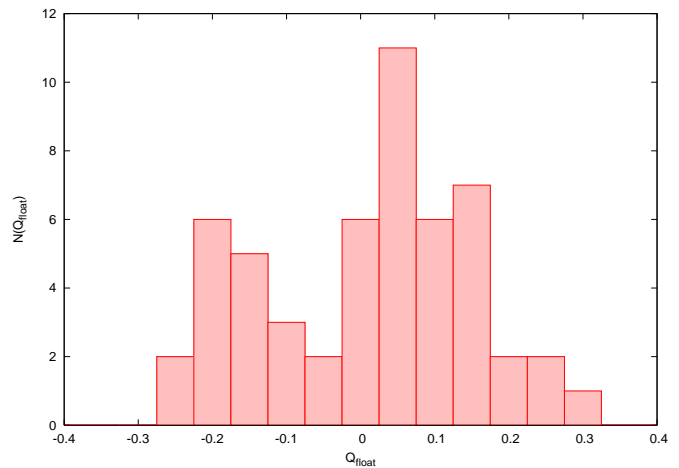,width=0.5\textwidth,silent=,angle=-90,clip=}}
\caption{Distribution of $Q_{float} - Q$, Eqs.~(\ref{Q-lat}), (\ref{Q}), calculated
on 55 configurations thermalized at $\beta=2.40$ on $16^4$ lattice.}
\label{fig:top_16_histo}
\end{figure}

The disadvantage of the algorithm is that the topological 
charge is not integer valued. However, for small enough accuracy 
of calculation at each simplex
the problem is likely to be inessential. In order to check this we performed
the following tests. First, we generated 500 sets of six random 5-dimensional
unit vectors which are geometrically assigned to the vertices of 5-dimensional tetrahedron,
the boundary of which consists of six 4-dimensional simplices and 
is $S^4$ topologically. Each generated set is the mapping $S^4\to S^4$
and we applied our algorithm to calculate the distribution of $Q_{float}$.
The resulting histogram is presented on Fig.~\ref{fig:top_simplex_histo}
and clearly shows that for this test the chosen accuracy is more then enough
since the topological charge is sharply peaked around integer values.
However, this simple test could not be completely convincing since
in Eq.~(\ref{Q-lat}) the uncertainties coming from each simplex ${\cal T}$ are accumulating.
Thus the deviation of the topological charge (\ref{Q-lat}) from nearest integer increases
with lattice volume $V$ and for given accuracy per each simplex the distribution
of the topological charge flattens in the limit $V\to\infty$.
Therefore we must ensure that for the lattices we're going to consider and for
the chosen accuracy per each simplex 
the distribution of $Q_{float}$ is indeed peaked around
integers. Below we discuss the physically relevant results obtained
on $16^4$ lattices at $\beta=2.4, 2.475$. Since the distribution of $Q_{float}$
is technical rather then physical issue it is presented in this section.
The distribution of non-integer part of the topological charge (\ref{Q-lat})
calculated on $16^4$ lattice at $\beta=2.40$ with 55 configurations is shown
on Fig.~\ref{fig:top_16_histo}. As expected the histogram is much broader
then that on Fig.~\ref{fig:top_simplex_histo}, but nevertheless is still reasonably sharp with
distribution width being $\approx 0.2$.  We conclude therefore that for the chosen parameters
and lattice volumes we could safely identify the topological charge as
\beq
\label{Q}
Q = \left[\,Q_{float}\,\right]\,,
\eeq
where $[x]$ denotes the nearest to $x$ integer number. We stress that
the identification (\ref{Q}) is not universal and is valid only for particular
range of parameters and lattice volumes. From now on the term ``topological charge''
will refer exclusively to the integer valued quantity $Q$.

Unfortunately, the presented algorithm turns out to be very time consuming.
As a matter of fact there is no hope to use it on single processor machines. Indeed,
the considered triangulation of one 4-dimensional hypercube consists of $16$ simplices and their
total number becomes prohibitively large for any reasonable lattice volumes.
However, it is very easy to implement the method on computer clusters since calculation
of the topological charge density is done independently for each simplex. 

\subsubsection{Testing the Algorithm}

\begin{figure}[t]
\centerline{\psfig{file=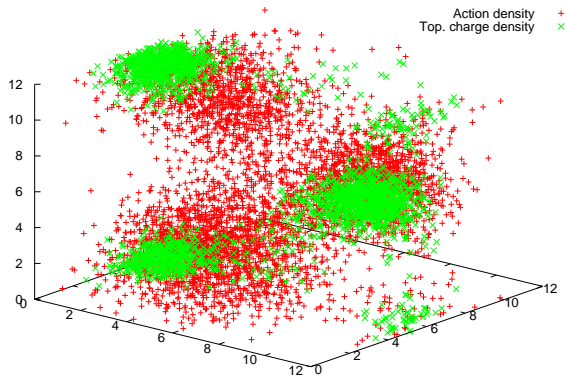,width=0.5\textwidth,silent=,angle=-90,clip=}}
\caption{Time slice of the action and the topological charge densities (\ref{q-lat})
on cooled $12^4$ configuration containing three instantons.}
\label{fig:testing-ins0}
\end{figure}

\begin{figure}[t]
\centerline{\psfig{file=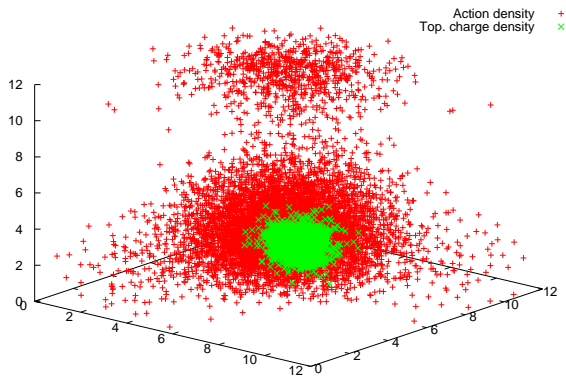,width=0.5\textwidth,silent=,angle=-90,clip=}}
\caption{Time slice of the action and the topological charge densities (\ref{q-lat})
on cooled $12^4$ configuration containing one anti-instanton.}
\label{fig:testing-ins1}
\end{figure}

The approximate integer valuedness of the topological charge is not the only test
which our algorithm should pass. At least we should convince ourselves that the method
works in the case of (quasi)classical configurations and compare it
with known topological charge constructions.

As far as the quasi-classical fields are concerned we applied our method to the same
set of highly cooled $12^4$ configurations generated in section~\ref{HPN-embedding}.
We have seen that the approximation quality of SU(2) gauge matrices
by the corresponding $\HP{1}$ $\sigma$-model fields was almost excellent (about 1\%).
Now we compare the local gauge action and the density of the topological charge on these
configurations. It turns out that they agree just nicely as shown on Fig.~\ref{fig:testing-ins0},
\ref{fig:testing-ins1}, where for a particular time slice $t=2$  the amount of red (green) points per 3-dimensional lattice
cube is proportional to the action (absolute value of the topological charge) density.
On the first configuration (Fig.~\ref{fig:testing-ins0}) our algorithm identified three instantons
with total topological charge being $Q_{float} = 2.91$.
On the second configuration (Fig.~\ref{fig:testing-ins1}) we found one anti-instanton and the topological charge is
$Q_{float} = -0.98$. Note that the regions where the topological charge
density is significant are rather large, their volumes are about $3^4$ in lattice units.

To summarize, we are confident that our method works as expected on quasi-classical
configurations. Next we would like to compare it with becoming standard nowadays
overlap-based topological charge definition~\cite{overlap}.

\subsubsection{Comparison with Overlap-Based Topological Charge}
\label{overlap-compare}

\begin{figure}[t]
\centerline{\psfig{file=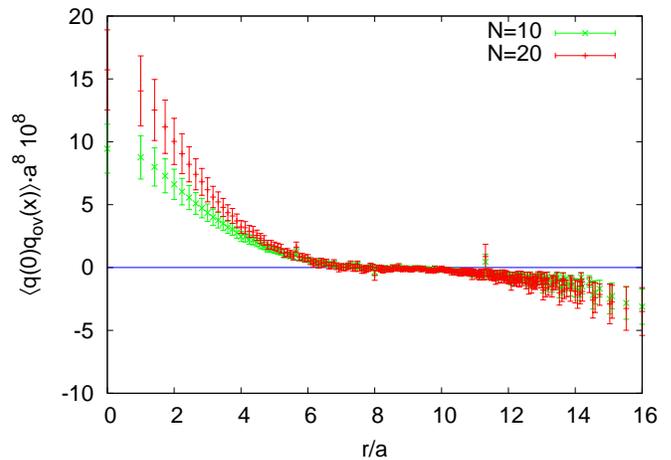,width=0.5\textwidth,silent=,angle=-90,clip=}}
\caption{Correlation function (\ref{overlap-3}) of the overlap- and
$\HP{1}$ $\sigma$-model based definitions of the topological charge density, see text.}
\label{fig:overlap}
\end{figure}

The validity of the topological charge construction we discussed so far could
only be convincing provided that we confront it with other known topological charge
definitions. As far as field-theoretical~\cite{DiVecchia:1981qi}
and geometrical~\cite{Luscher,Stone} constructions are concerned, 
they could be applied in fact only on cooled configurations~\cite{Pugh}
and we refuse the corresponding comparison for this reason. On the other hand,
the definition~\cite{overlap} based on the overlap Dirac operator~\cite{Neuberger:1997fp}
\beq
\label{overlap-1}
q_o(x) = -\mathrm{tr}\,\gamma^5\left( 1 - \frac{1}{2} D_{x,x}\right)
\eeq
is free from the above disadvantages.
For reasons to be discussed in section~\ref{qq-corr} we will not consider
the full trace in Eq.~(\ref{overlap-1}), but restrict it to $N$ lowest eigenmodes
$D \psi^{(i)} = \lambda_i \psi^{(i)}$
\beqn
\label{overlap-2}
q^N_o(x) = - \sum\limits_{i < N} (1 - \frac{\lambda_i}{2}) c^{(i)}(x)\,, \\
c^{(i)}(x) = {\psi^{(i)}_x}^\dagger \gamma^5 \psi^{(i)}_x\,. \nonumber
\eeqn
The quantity $q^N_o$ is known as the effective topological density~\cite{Horvath-qq}.

In order to make a quantitative comparison of the definitions (\ref{q-lat}) and (\ref{overlap-2})
we measured the correlation function~\cite{Cundy:2002hv}
\beq
\label{overlap-3}
C(x) = \langle q(0) \, q^N_o(x) \rangle
\eeq
on 18 statistically independent $16^4$ configurations thermalized at $\beta=2.475$.
The results corresponding to $N=10$ and $N=20$ lowest eigenmodes taken into account
are presented on Fig.~\ref{fig:overlap}. It is apparent that both definitions
are indeed compatible with each other and determine the topological density consistently.
Moreover, the graph shown on Fig.~\ref{fig:overlap} is very similar to the corresponding
one given in Ref.~\cite{Cundy:2002hv}, where the overlap-based and field-theoretical
constructions were compared.

To summarize, we found clear evidences that our construction of the topological charge
density is locally correlated with overlap-based definition. This provides
a stringent test of our approach and allows us move directly to the investigation
of the SU(2) gauge fields topology.

\section{Topology of SU(2) Gauge Fields}
\label{HPN-topology}

\begin{table}
\centerline{
\begin{tabular}{|p{0.05\textwidth} | p{0.08\textwidth} | p{0.05\textwidth} |
                p{0.05\textwidth} | p{0.08\textwidth} | p{0.05\textwidth} |}
\hline
$\beta$ & $a$,fm & $L_s$ & $L_t$ & $V^{phys}, \mathrm{fm}^4$ & $ N_{conf}$ \\ \hline
2.40  & 0.1193(9) &  16 & 16 & 13.3(4) & 55 \\
2.475  & 0.0913(6) &  16 & 16 & 4.6(1) & 18 \\ \hline
\end{tabular}
}
\caption{Simulation parameters.}
\label{tab:params}
\end{table}

In this section we discuss the measurements performed with $\HP{1}$ $\sigma$-model
construction of the topological charge in pure SU(2) lattice gauge theory.
In section~\ref{HPN-dynamics} the dynamics of embedded $\HP{1}$ $\sigma$-model
is briefly considered. Then in section~\ref{global-Q} we present our results concerning
global topological charge distribution and topological susceptibility.
Section~\ref{top-density} is devoted to the detailed investigation of the topological charge
density bulk distribution and the corresponding correlation function.
Our calculations were performed on two sets (Table~\ref{tab:params}) of statistically independent
SU(2) gauge configurations generated with standard Wilson action. The lattice spacing
values quoted in the Table~\ref{tab:params} are taken from Ref.~\cite{Lucini:2001ej} and fixed by the physical
value of SU(2) string tension $\sqrt{\sigma} = 440~\mathrm{MeV}$. As we noted already
the method we employed for topological charge density calculation is very time consuming and this
explains the relatively small number of configurations we were able to analyze.

\subsection{Dynamics of the Embedded $\HP{1}$ $\sigma$-model}
\label{HPN-dynamics}

\begin{figure}[t]
\centerline{\psfig{file=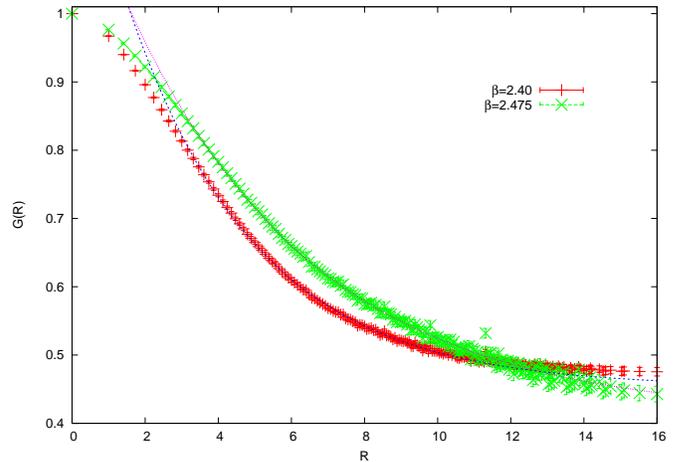,width=0.5\textwidth,silent=,angle=-90,clip=}}
\caption{The correlation function (\ref{spin-spin}) in the embedded
$\HP{1}$ $\sigma$-model. Lines correspond to the best fitting curves (\ref{spin-fit}).}
\label{fig:scalar}
\end{figure}

Given that in our approach the calculation of the topological properties of SU(2) gauge
fields requires the construction of $\HP{1}$ $\sigma$-model it is natural to consider
the dynamics of the embedded scalar fields. The basic correlation function in
the $\HP{n}$ $\sigma$-models is given by
\beq
\label{spin-spin}
G(x) ~=~ \langle \mathrm{Sc} \, \mathrm{tr} \, [\, P(0) P(x) \,] \rangle\,,
\eeq
where $P_{ik}$ are the quaternionic valued projector matrices (\ref{projector}).
Note that we didn't subtracted the disconnected part of $G(x)$.
The physical meaning of the correlation function (\ref{spin-spin}) could well be illustrated
for $\HP{1}$ $\sigma$-model where it reduces up to irrelevant constant contributions to
$G(x) = \langle \sum_A n^A_0 n^A_x \rangle$. On general grounds one expects that the embedded
$\HP{n}$ $\sigma$-models could be in two different phases distinguished
by $G(x)$. Namely, the correlator (\ref{spin-spin}) falls off exponentially at large
$|x|$ in massive (disordered) phase and only with some inverse power of $|x|$ in the massless
(ordered) phase. Taking into account the quality of the approximation (\ref{approx1})
discussed in section~\ref{HPN-embedding} one could argue that the confinement (deconfinement)
phase of gluodynamics might be reflected by the disordered (ordered) phase in the corresponding
$\sigma$-models.  At present it is too early to discuss this conjecture quantitatively, but
our preliminary studies indicate that it is indeed reasonable.
The data we have right now refer to the confinement phase exclusively where we expect
the exponential behavior of the correlator (\ref{spin-spin})
\beq
\label{spin-fit}
G( |x|\to\infty )  ~\approx~ C_0 \,\, e^{-m |x|} ~+~ C_1\,.
\eeq

\begin{figure}[t]
\centerline{\psfig{file=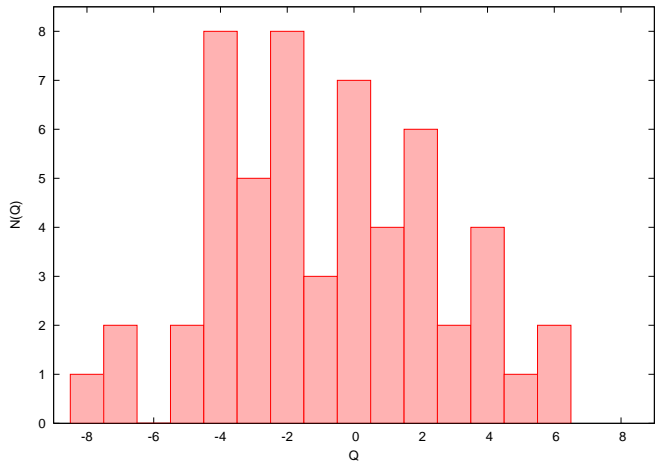,width=0.5\textwidth,silent=,angle=-90,clip=}}
\caption{Topological charge distribution calculated on $\beta=2.40$ configurations.}
\label{fig:Q_histo}
\end{figure}

The results of our measurements of the correlation function (\ref{spin-spin})
are presented on Fig.~\ref{fig:scalar} (left). To investigate the large distance behavior
of $G(x)$ we fitted the data to Eq.~(\ref{spin-fit}) in various ranges of $R=|x|$.
The fits turn out to be rather stable,
the best fitting curves are shown on Fig.~\ref{fig:scalar} by dashed lines.
The results of various fits strongly suggest that the parameter $m$ entering Eq.~(\ref{spin-fit})
is nothing but the SU(2) string tension
\beq
\label{spin-m}
m ~=~ a \, \sqrt{\sigma}\,,
\eeq
being equal to $m_{\beta=2.40} = 0.28(2)$ and $m_{\beta=2.475} = 0.19(1)$ at the $\beta$ values
considered. In fact, Eq.~(\ref{spin-m}) indirectly supports the above
made conjecture on the phase diagram of the embedded $\HP{1}$ $\sigma$-model. 

\subsection{Topological Charge}
\label{global-Q}

We should mention that the limited set of configuration we have in our disposal
prevents us from studying the topological charge distribution and
related quantities with sufficient accuracy. Nevertheless our results
allow to draw some preliminary conclusions about the method we presented.
There are two basic quantities to be discussed here, the topological charge
(\ref{Q}) distribution and the topological susceptibility $\chi = \langle Q^2 \rangle/V$.

The distribution of the topological charge for $\beta=2.40$ is shown on
Fig.~\ref{fig:Q_histo}. Although the width of the histogram is roughly what is expected at this
$\beta$ value, the shape of the distribution seems to indicate the lack of statistics.
In particular, it is hardly possible to fit it with Gaussian profile. Moreover,
the noticeable asymmetry in odd/even $Q$ values should also be mentioned, which
probably hints on some subtle issues in our topological charge definition.
Anyhow, the evident lack of statistics
prohibits us to make solid conclusions at the moment. For the same reason
we do not present the analogous histogram at $\beta=2.475$ since the statistics
in this case is even worse.

As far as the $\langle Q^2 \rangle$ and the topological susceptibility are concerned,
their values turn out to be
\beq
\label{Q2}
\begin{array}{c|c|c}
\beta \quad & \,\,\,\,\langle Q^2 \rangle \quad & \,\,\,\, \chi \\ \hline
2.40  \quad & \,\,\,\, 10.8 \pm 1.8       \quad & \,\,\,\,(190(10) \mathrm{~MeV})^4  \\ \hline
2.475 \quad & \,\,\,\,  3.9 \pm 1.0       \quad & \,\,\,\,(192(12) \mathrm{~MeV})^4
\end{array}
\eeq
which indicate a perfect scaling and are fairly consistent with data of Ref.~\cite{Lucini:2001ej}.

To summarize, at present our calculation of the global topological charge is plagued by the lack
of statistics and should be improved to quantify our method. Nevertheless,
qualitatively our results are in agreement with the literature and don't
show any pathology in the approach we have developed.

\subsection{Topological Charge Density}
\label{top-density}

\begin{figure}[t]
\centerline{\psfig{file=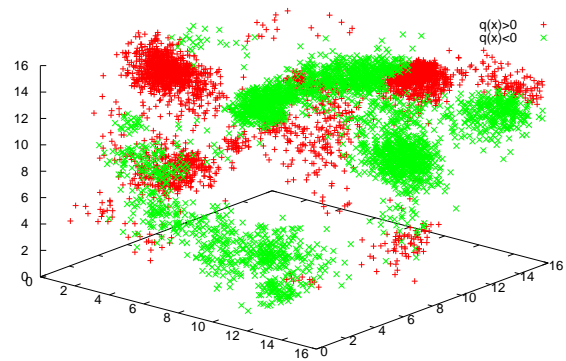,width=0.5\textwidth,silent=,angle=-90,clip=}}
\caption{Visualization of the topological charge density distribution
in a particular time slice of $16^4$, $\beta=2.40$ thermalized configuration.}
\label{fig:q2}
\end{figure}

The topological charge density $q(x)$ is defined by Eq.~(\ref{q-lat}) and 
provides quantitative characterization the topological charge bulk distribution.
As a first qualitative illustration we present visualization of $q(x)$ in particular
time-slice of four dimensional lattice. It turns out that the corresponding pictures
at various slices look very similar. The typical distribution of $q(x)$
is shown on Fig.~\ref{fig:q2} (data is taken on $16^4$, $\beta=2.40$ lattices).
Here the density of points is proportional to the absolute value $|q(x)|$ while red (green)
points mark the regions of positive (negative) topological charge density. Note that the
empty areas correspond to indeed either vanishing or at least utterly small $|q(x)|$.

Several qualitative comments are now in order. It is apparent and remarkable that the topological
charge density is distributed in localized regions which roughly look like four dimensional balls.
As a matter of fact this regions are almost spherically symmetric resembling instantons
from the first sight. We are in hustle to add however that we don't claim that these
regions are indeed the instantons (see below).
In either case the lumpy structure of the topological charge density is present in
all configurations we have and  could be taken as firmly established.
Note that this result is not new, the lumpy distribution of the topological charge
in ``hot'' (not cooled, smoothed or smeared in any way) vacuum configurations
was observed already in Refs.~\cite{Gattringer,DeGrand,Hip,Horvath:2002gk}.
However, we found the lumps completely independently
using newly developed method which not only confirms their existence but also
provides new opportunities to investigate the structure of the lumps.
The second point is that the lumps $\cL_i$ in $q(x)$ distribution are significantly dilute,
their total number in every time slice could be estimated as $\lesssim 10$, the corresponding
lumps density being $D_\cL \approx 3 \cdot 10^{-3}$ in lattice units which translates
to $D_\cL\approx 10 \mathrm{~fm}^{-4}$ for $\beta=2.40$.
At the same time it is apparent that the lumps are clearly separated from each other and
their characteristic size is of order
$\rho_\cL \approx (1\div 2) a \approx 0.12\div 0.24 \mathrm{~fm}$.

The reminder of this section is organized as follows. In section~\ref{lumps}
we describe in details our approach aimed to investigate the lumpy structure described above.
In particular, we argue that in order to quantitatively define the notion of lumps
it is mandatory to introduce the physically motivated cutoff $\Lambda_q$ on $|q(x)|$ values
and to ensure that final results are cutoff independent in reasonably large range of $\Lambda_q$.
In section~\ref{limit}
we explore the limit $\Lambda_q \to 0$ which seems to be unphysical for us, but
in which we find that the topological charge organizes into some global structures
reminiscent to ones discovered recently~\cite{Horvath-structures}.
Then in section~\ref{qq-corr} we present our results for the topological charge density correlation function
and discuss some related issues. Until section~\ref{scaling} we consider only
the data collected on $16^4$, $\beta=2.40$ lattices, while in section~\ref{scaling}
the comparison is made with the results obtained at $\beta=2.475$.

\subsubsection{Lumps in $q(x)$ Distribution}
\label{lumps}

In order to investigate quantitatively the nature of the lumps $\cL_i$ in the topological
charge density we have to define first an algorithm to identify the lumps. The natural procedure is to apply
some cutoff $\Lambda_q$ to the topological density and to treat the small values $|q(x)|< \Lambda_q$
as identical zero.  It is expected that for reasonably wide range of $\Lambda_q$ the lumps $\cL_i$
would look like an isolated 4-dimensional regions of sign-coherent topological charge.
It is clear that the number of lumps and other their characteristics
depend upon the cutoff $\Lambda_q$. We expect that the universal physical properties
of the lumps in $q(x)$ distribution should be independent or depend only mildly on $\Lambda_q$
(in the second case the bias introduced by $\Lambda_q$ dependence is to be included into systematic errors).

Notice the following two extreme cases. Evidently, for large $\Lambda_q$ there would be no
lumps at all and hence the maximal cutoff coincides by order of magnitude  with 
the maximal value of $|q(x)|$. On the other hand for $\Lambda_q = 0$
one expects to find a large amount of very small lumps consisting of just one lattice point.
There are at least two reasons to justify this assumption. First,
this is because we constructed the $\HP{1}$ $\sigma$-model starting from ``hot''
thermalized gauge configurations and $\ket{q}$ fields reflect partially
the ultraviolet noise present in the original gauge potentials.
Therefore, away from the lumps the topological density is expected
to fluctuate widely around zero while being almost vanishing in magnitude.
Secondly, it is quite evident that it makes no sense to assign any physical meaning
to utterly small $|q(x)|$. The physically meaningful minimal $\Lambda_q$ value could be estimated
by order of magnitude as the ratio of typical topological charge and the lattice volume, which
for our lattices is $\approx 10^{-5}$. We conclude therefore that the extremely small cutoff
$\Lambda_q \cdot a^4 < 10^{-5}$ could not have any physical significance and the universal properties
of the topological density bulk distribution (universality window)
should be looked for in $\Lambda_q \cdot a^4 > 10^{-5}$ region.

\begin{figure}[t]
\centerline{\psfig{file=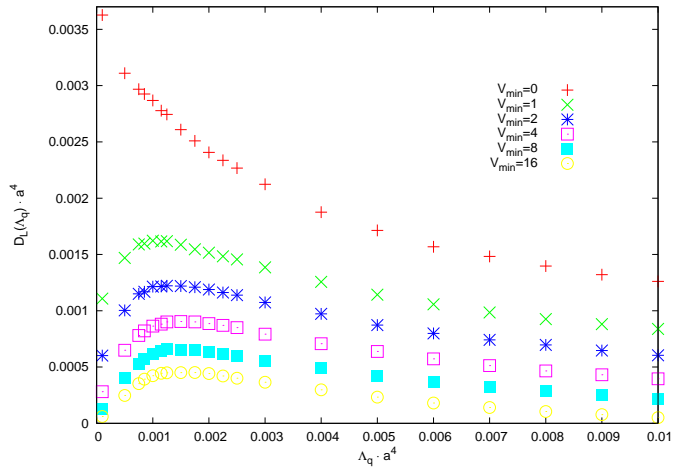,width=0.5\textwidth,silent=,angle=-90}}
\caption{Dependence of the lumps density $D_\cL$ upon the cutoff $\Lambda_q$
for various values of $V_{min}$, Eq.~(\ref{volume-cut}).}
\label{fig:lumps-number-1}
\end{figure}

\begin{figure}[t]
\centerline{\psfig{file=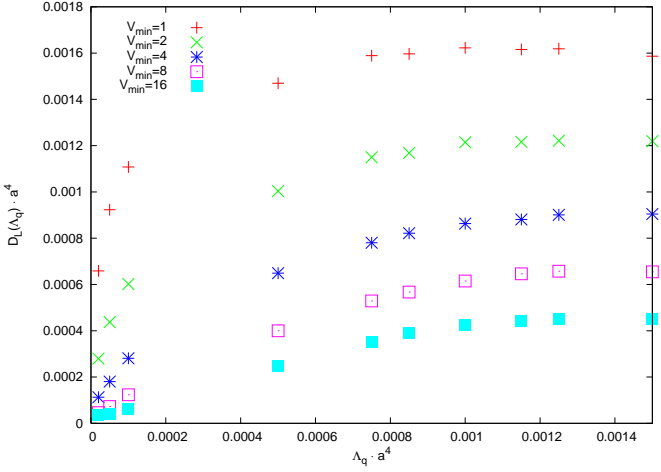,width=0.5\textwidth,silent=,angle=-90}}
\caption{$D_\cL(\Lambda_q)$ dependence in the range $10^{-5} < \Lambda_q \cdot a^4 < 1.5 \cdot 10^{-3}$
for various $V_{min}$, Eq.~(\ref{volume-cut}).}
\label{fig:lumps-number-2}
\end{figure}

Let us consider first the dependence of the total lumps density $D_\cL$ upon the cutoff $\Lambda_q$,
which is shown by the upper curve on Fig.~\ref{fig:lumps-number-1}.
As expected from the discussion above $D_\cL$ is monotonically decreasing function of $\Lambda_q$
and diverges in the limit $\Lambda_q \to 0$.
However, the dependence $D_\cL(\Lambda_q)$ becomes less trivial if we consider only lumps satisfying
\beq
\label{volume-cut}
V( \cL_i ) ~ > ~ V_{min} \gtrsim 1 \,,
\eeq
where $V( \cL_i )$ is the total number of sites constituting the lump $\cL_i$
and $V_{min}$ is taken to be of order few units. In fact, the constraint (\ref{volume-cut})
is quite natural since for $V_{min}=1$ it cuts only smallest fraction of UV-scale lumps. It is remarkable that 
already from $V_{min}=1$ onward the shape of $D_\cL(\Lambda_q)$ dependence is universal,
the only thing which distinguishes the curves at various $V_{min}$ is the scale
of the density $D_\cL(\Lambda_q)$. Note that the behavior of $D_\cL$
in the limit $\Lambda_q \to 0$ is changing qualitatively once the volume cutoff is imposed.
As is apparent from Fig.~\ref{fig:lumps-number-2}, for any $V_{min}>0$ we have
\beq
\lim\limits_{\Lambda_q \to 0} \left.  D_\cL \,\right|_{V_{min} > 0} = const < \infty
\eeq
and this is discussed in the next section.
The maximum of the function $D_\cL(\Lambda_q)$ at $V_{min} > 0$ and its rather mild
dependence on $\Lambda_q$ for
\beq
\label{Lambda-low}
\Lambda_q > 10^{-3} \cdot a^{-4} = (290\mathrm{~MeV})^4
\eeq
suggests that $\Lambda_q$ universality window starts from this value.

\begin{figure}[t]
\centerline{\psfig{file=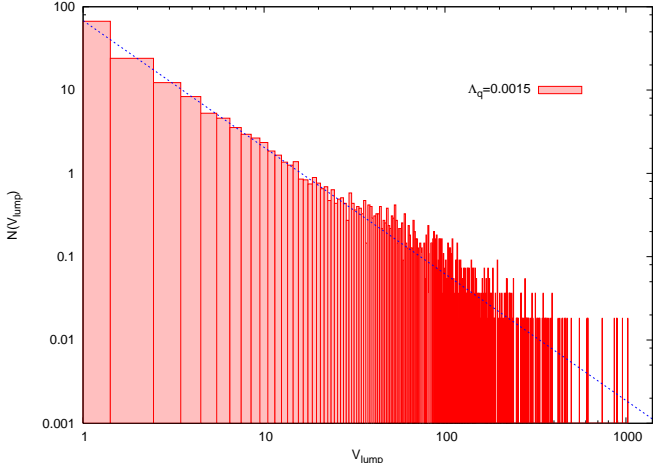,width=0.5\textwidth,silent=,angle=-90,clip=}}
\caption{Distribution of lumps volumes, Eq.~(\ref{VQ-defs}),
for $\Lambda_q \cdot a^4 = 1.5 \cdot 10^{-3}$ calculated on $16^4$, $\beta=2.40$
lattices and normalized by the number of configurations. Solid line is the fit to Eq.~(\ref{VQ-fit}).}
\label{fig:V-0015-histo}
\end{figure}

\begin{figure}[t]
\centerline{\psfig{file=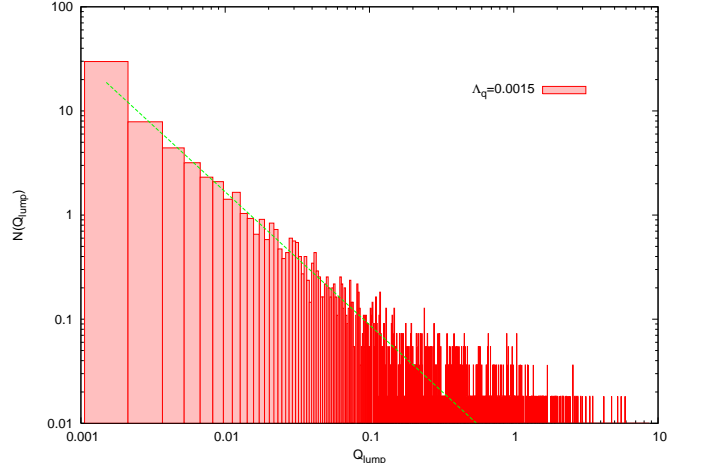,width=0.5\textwidth,silent=,angle=-90,clip=}}
\caption{Lumps charge, Eq.~(\ref{VQ-defs}), distribution for
$\Lambda_q \cdot a^4 = 1.5 \cdot 10^{-3}$ calculated on $16^4$, $\beta=2.40$ lattices and
normalized by the number of configurations. Solid line is the fit to Eq.~(\ref{VQ-fit}).}
\label{fig:Q-0015-histo}
\end{figure}

Let us now consider the volumes and the topological charges of the lumps $\cL_i$
\beq
\label{VQ-defs}
V(\cL_i) = \sum\limits_{ x \in \cL_i} 1\,, \qquad
Q(\cL_i) = \sum\limits_{ x \in \cL_i} q(x)\,,
\eeq
identified for given value of $\Lambda_q$. The distributions of $V(\cL_i)$ and
$Q(\cL_i)$ for $\Lambda_q \cdot a^4 = 1.5 \cdot 10^{-3}$ calculated on $16^4$, $\beta=2.40$ lattices
and normalized by the number of configurations are shown on Fig.~\ref{fig:V-0015-histo}
and Fig.~\ref{fig:Q-0015-histo} respectively.
A few points are worth to be mentioned here. First, we see that both volume and charge
distributions are well described by power laws
\beqn
\label{VQ-fit}
N[ V(\cL_i) ] = c_V \cdot [V(\cL_i)]^{-\alpha_V}\,, \\
N[ Q(\cL_i) ] = c_Q \cdot [Q(\cL_i)]^{-\alpha_Q}\,, \nonumber
\eeqn
for not exceptionally large $V(\cL_i)$ and $Q(\cL_i)$. On the other hand the appearance
of extremely large lumps with volumes $V(\cL_i) \gtrsim 100 \cdot a^4$ and charges $Q(\cL_i) \gtrsim 1$ 
is cutoff dependent (see below). Therefore, let us concentrate on the behavior
(\ref{VQ-fit}) valid for moderate $V(\cL_i)$ and $Q(\cL_i)$. 
It turns out that both histograms in the almost entire ranges of volumes and charges
are well described by Eq.~(\ref{VQ-fit}) and the fitted values of power exponents
are given by
\beqn
\left. \alpha_V \right|_{\Lambda_q\cdot a^4 = 0.0015} & = & 1.52(8)\,, \\
\left. \alpha_Q \right|_{\Lambda_q\cdot a^4 = 0.0015} & = & 1.32(20)\,, \nonumber
\eeqn
where a conservative error estimates are quoted. Apparently
both power exponents are compatible with each other and
it is tempting to conclude that for $\Lambda_q \cdot a^4 = 1.5 \cdot 10^{-3}$
they are equal to
\beq
\label{VQ-fixed}
\alpha_V ~=~ \alpha_Q ~=~ 3/2\,.
\eeq

\begin{figure}[t]
\centerline{\psfig{file=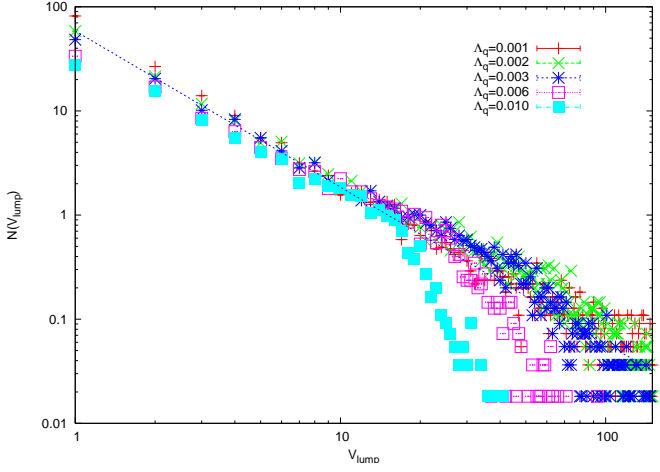,width=0.5\textwidth,silent=,angle=-90,clip=}}
\caption{Lumps volume distribution for various $\Lambda_q$ given in lattice units.
Solid line is the fit (\ref{VQ-fit}) with $\alpha_V = 3/2$.}
\label{fig:V-histo}
\end{figure}

\begin{figure}[t]
\centerline{\psfig{file=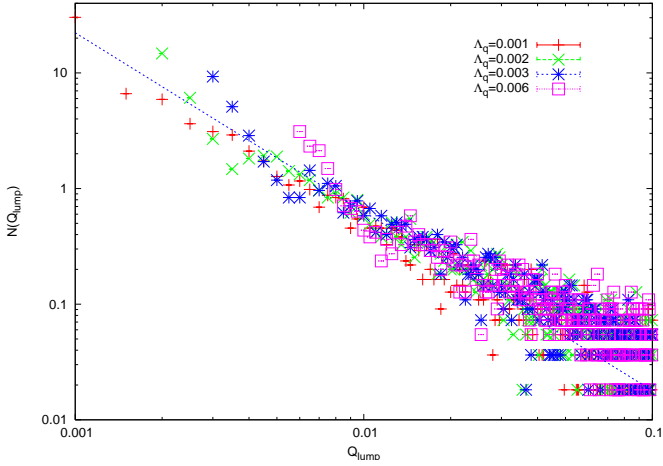,width=0.5\textwidth,silent=,angle=-90,clip=}}
\caption{Distribution of lumps charges for different $\Lambda_q$ given in lattice units.
Solid line is the fit (\ref{VQ-fit}) with $\alpha_Q = 3/2$.}
\label{fig:Q-histo}
\end{figure}

\noindent
Clearly, we should check this conjecture at very least by investigating the dependence
of power exponents upon the cutoff $\Lambda_q$. The corresponding histograms are presented
on Fig.~\ref{fig:V-histo} and  Fig.~\ref{fig:Q-histo} where we have collected
the distributions of $V(\cL_i)$ and $Q(\cL_i)$
obtained for various $\Lambda_q$ on $16^4$, $\beta=2.40$ configurations. The solid lines on the figures represent
the one-parameter fits to Eq.~(\ref{VQ-fit}) in which we have fixed $\alpha_V$ and $\alpha_Q$
according to Eq.~(\ref{VQ-fixed}). It is remarkable that the volumes of the lumps
are distributed well within numerical uncertainties in agreement with (\ref{VQ-fit}) 
at all values of the cutoff. The only
deviation from (\ref{VQ-fit}) is seen at large $\Lambda_q$, $V(\cL_i)$ and could be easily
understood. Indeed, for very large $\Lambda_q$ we will not find 
any lumps at all and with diminishing cutoff we start seeing lumps from some finite volume only.
This could explain the apparent drop in the $V(\cL_i)$ distribution for
$\Lambda_q \cdot a^4$ equal to $6 \cdot 10^{-3}$ and $10^{-2}$. Essentially the same arguments apply
to the $Q(\cL_i)$ histograms which show a slight deviation from (\ref{VQ-fit})
for large $\Lambda_q$ and small $Q(\cL_i)$.
Thus the have found an universal (cutoff independent) behavior 
(\ref{VQ-fit}), (\ref{VQ-fixed}) of the lumps volume and charge distributions for
$\Lambda_q \cdot a^4 > 10^{-3}$. In the next section we show that for smaller
$\Lambda_q$ this universality does not hold. Therefore we have clear evidences
that the lower bound of the cutoff universality window is indeed given by (\ref{Lambda-low}).

Let us make a few comments concerning the distribution (\ref{VQ-fit}), (\ref{VQ-fixed}).
The simplest possible picture behind the lumpy structure of the topological
density is that $q(x)$ is non-vanishing only within the four dimensional balls
of radius $\rho$.  It is reminiscent to the well known instantonic picture
of the vacuum and in this case the power law (\ref{VQ-fit}), (\ref{VQ-fixed}) translates into
\beq
N[\rho] \,d \rho  ~ \sim ~ \frac{d\rho}{\rho^3}\,,
\eeq
which is precisely the instantons distribution elaborated in Ref.~\cite{Diakonov}.
While the microscopic view of the lumps as four dimensional balls of sign-coherent
topological charge looks quite natural, it is in fact incompatible
with instantons. Indeed, Eqs.~(\ref{VQ-fit}), (\ref{VQ-fixed}) imply that the lumps
are objects with fractional and arbitrary small
charge. As far as the large lumps with $Q(\cL_i) \approx 1$ are concerned
we discuss them in the next section, but right now it is enough to mention
that their number is negligible compared to the total number of lumps.
The conclusion is that the majority of the lumps are not the instantons.

\begin{figure}[t]
\centerline{\psfig{file=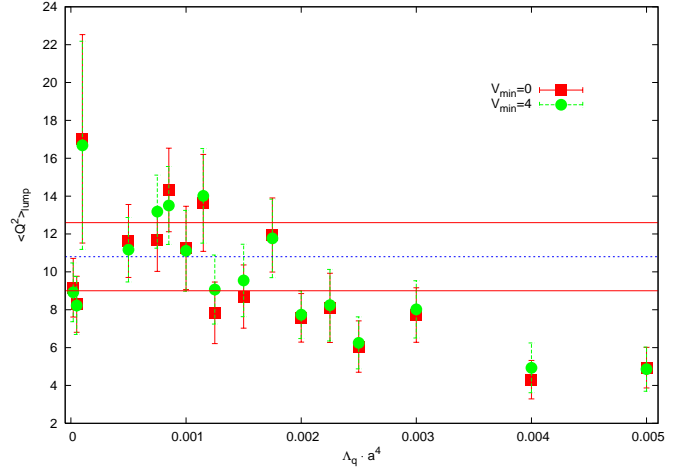,width=0.5\textwidth,silent=,angle=-90}}
\caption{Lumps limited squared topological charge $\langle Q^2 \rangle_{lump}$ versus
cutoff $\Lambda_q$. Lines represent $\langle Q^2 \rangle$, Eq.~(\ref{Q2}), and
the corresponding error bounds.}
\label{fig:Q2-lumps}
\end{figure}

The question of primary importance is whether the total topological charge
is dominated by the lumps. The relevant quantity is
the lumps limited topological susceptibility which is essentially $\langle Q^2 \rangle$
measured on the lumps only. Note that this quantity
naturally bounds the physically acceptable cutoff from above. Indeed, for vanishing
cutoff the lumps limited average $\langle Q^2 \rangle_{lump}$
coincides with the full one, $\langle Q^2 \rangle_{lump} = \langle Q^2 \rangle$,
while at very large $\Lambda_q$ we would have $\langle Q^2 \rangle_{lump} = 0$.
The characteristic cutoff at which the topological susceptibility drops down is
definitely the upper bound on physical $\Lambda_q$ and {\it a priori} could be
on the either side of the value quoted in (\ref{Lambda-low}).
The lumps would be clearly unphysical if $\Lambda_q$ range happens to be empty.

We have measured the lumps limited squared topological charge $\langle Q^2 \rangle_{lump}$
on $16^4$ $\beta=2.40$ configurations, the result is presented on Fig.~\ref{fig:Q2-lumps}.
Note that we also checked the dependence of $\langle Q^2 \rangle_{lump}$ on the volume
cutoff (\ref{volume-cut}) which turns out to be trivial as is clear from the figure.
Indeed, the data points with volume cut $V_{min} = 4$ imposed are falling on the top of
the data with no volume cut at all.
It is apparent that the topological susceptibility is indeed lumps dominated provided that 
\beq
\label{Lambda-high}
\Lambda_q < 2 \cdot 10^{-3} \, a^{-4} = (350\mathrm{~MeV})^4\,.
\eeq
Note that the universality window (\ref{Lambda-low}), (\ref{Lambda-high}) appears
to be rather small. However, this does not appear completely unexpected. Indeed, our original intent
was to find a plateau in the cutoff dependence of various quantities characterizing the lumps,
at which their first $\Lambda_q$ derivative vanishes. In particular, the universality window
(\ref{Lambda-low}), (\ref{Lambda-high}) could be estimated rather precisely by noting that
the maximum of the lumps density $D_\cL(\Lambda_q)$ at any $V_{min} \gtrsim 1$ 
is located roughly at
\beq
\label{Lambda-ref}
\Lambda^*_q ~=~ (1 \div 1.5) \, 10^{-3} \cdot a^{-4} ~=~ (314(16)\mathrm{~MeV})^4\,,
\eeq
which almost coincides with the universality window range.

Finally, let us note that the question of dislocations, which {\it a priori}
could be crucial for our investigation, is likely to be irrelevant in fact.
For general arguments, which we could just verbosely repeat here
and which show that dislocations are not essential in our case, the reader
is referred to Ref.~\cite{Horvath:2002gk}. We could only add that the dislocation
filtering (as well as partial filtering of the UV noise) is built into our approach.
Moreover, we don't see any sign of dislocations in the topological susceptibility
which seems to be perfectly scaling and is in a good agreement with its accepted value.
Following Ref.~\cite{Horvath:2002gk} we stress that the notion of dislocation is indispensable from
both the gauge action and the topological charge operator. Therefore due to the specific
properties of our topological charge construction we expect that the issue
of dislocations inherent to Wilson gauge action is not applicable in our case.

\subsubsection{Exploring the Limit $\Lambda_q \to 0$}
\label{limit}

In this section we explore the limit of vanishing cutoff imposed on the topological charge density.
Although it is not clear for us what is the physical relevance of the limit $\Lambda_q\to 0$
it nevertheless seems to be interesting to consider in connection with recent observation
of low-dimensional global structures
in the topological charge density distribution~\cite{Horvath-structures}.
Essentially, these global structures are defined as topological charge sign-coherent regions
and thus are similar to the lumps we're investigating.

In fact, the non-trivial features of the $\Lambda_q \to 0$ limit could be seen already
on Fig.~\ref{fig:lumps-number-2} from which it follows that once the mildest
$V_{min}=1$ volume cut is imposed the density of lumps rapidly diminishes with decreasing
$\Lambda_q$ provided that $\Lambda_q \cdot a^4 \lesssim 10^{-3}$. The dependence $D_\cL(\Lambda_q)$
in the range (see Fig.~\ref{fig:lumps-number-2})
\beqn
\label{small-range}
\Lambda_q \,\in\, & [\,2 \cdot 10^{-5} \cdot a^{-4} \,;\, 10^{-3} \cdot a^{-4} \,] & = \\
      =           & [\,(110\mathrm{~MeV})^4 \,;\, (290\mathrm{~MeV})^4\,] & \nonumber
\eeqn
makes it apparent that $D_\cL(0)$ is very small if not non-vanishing.
However, the total number of lumps (no volume cut imposed) is divergent,
hence the dominant contribution at small $\Lambda_q$ comes from lumps consisting
of just one point. Then the question is whether we have any other contributions to $D_\cL$
or everything is exhausted by one-point lumps.

\begin{figure}[t]
\centerline{\psfig{file=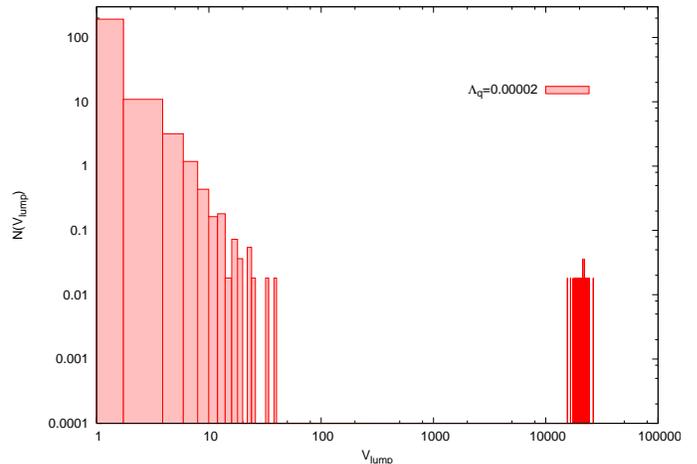,width=0.5\textwidth,silent=,angle=-90,clip=}}
\caption{Lumps volume distribution for fixed cutoff $\Lambda_q \cdot a^4 = 2 \cdot 10^{-5}$.}
\label{fig:V-histo-small}
\end{figure}

\begin{figure}[t]
\centerline{\psfig{file=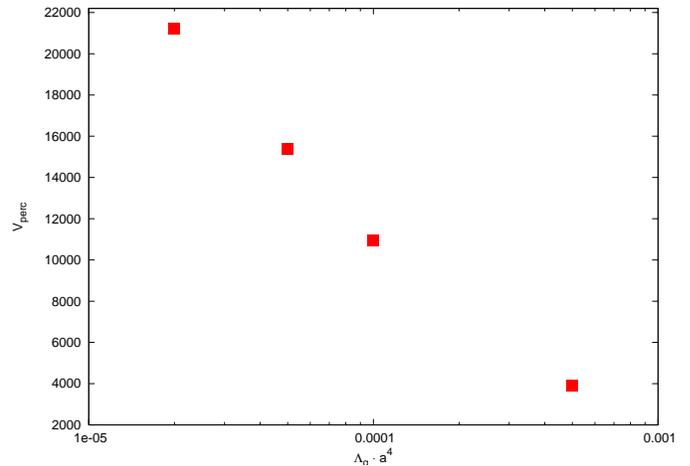,width=0.5\textwidth,silent=,angle=-90,clip=}}
\caption{Volumes of percolating lumps at various $\Lambda_q$.
Note that percolating lumps disappear at $\Lambda_q \cdot a^4 \approx 7 \cdot 10^{-4}$.}
\label{fig:V-perc-small}
\end{figure}

To investigate this issue we considered the lumps volume distribution at 
$\Lambda_q \cdot a^4 = 2 \cdot 10^{-5}$ and it turns out (see Fig.~\ref{fig:V-histo-small})
that it is qualitatively distinct
from the case of moderate cutoff. First, we clearly see that the histogram separates
into two disconnected pieces corresponding to small and extremely large (percolating) lumps.
As far as the percolating lumps are concerned we didn't investigated their properties
in details for reasons to be explained shortly. We only mention that there are almost always
only two percolating lumps on our configurations carrying an extremely large and opposite
topological charge. Qualitatively our results are in accord with that of
Refs.~\cite{Horvath-structures},
where the analogous percolating global structures were found.
If we now turn to the consideration of small lumps distribution,
it also follows sharply the law (\ref{VQ-fit}). However, the power exponent is drastically changing
\beq
\left. \alpha_V\right|_{\Lambda_q \cdot a^4 = 2\cdot 10^{-5}} = 2.9(1)\,.
\eeq

\begin{figure}[t]
\centerline{\psfig{file=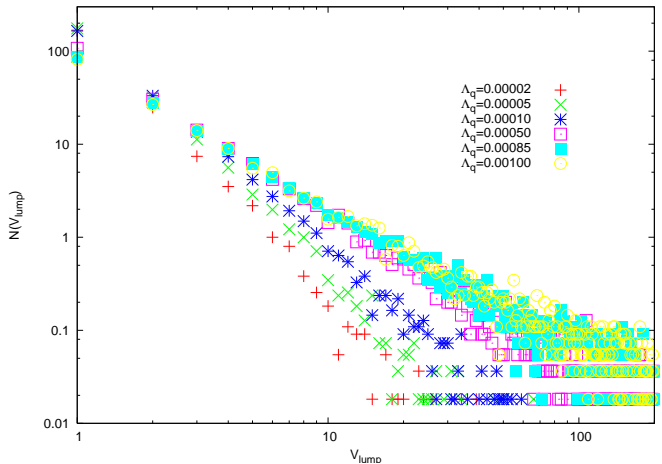,width=0.5\textwidth,silent=,angle=-90,clip=}}
\caption{Distribution of the small volume lumps at various $\Lambda_q$.
Note that it still follows the power law (\ref{VQ-fit}).}
\label{fig:alpha1}
\end{figure}

\begin{figure}[t]
\centerline{\psfig{file=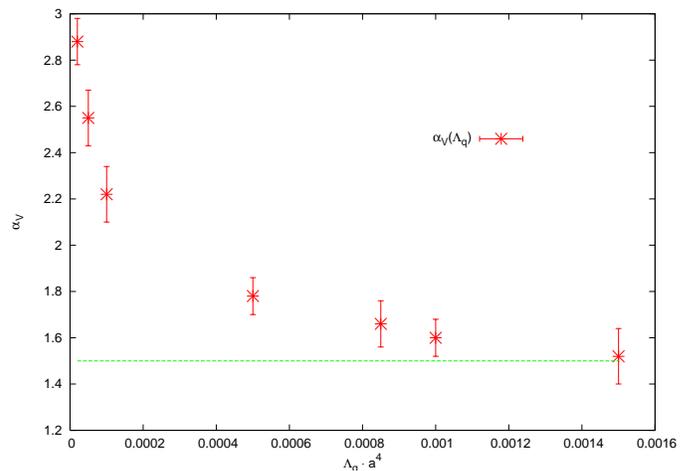,width=0.5\textwidth,silent=,angle=-90,clip=}}
\caption{The power exponent $\alpha_V$, Eq.~(\ref{VQ-fit}), as function of $\Lambda_q$
in the range (\ref{small-range}). Line represents the conjectured universal value (\ref{VQ-fixed}).}
\label{fig:alpha2}
\end{figure}

Next we compare the lumps volume distributions at various $\Lambda_q$
from the range (\ref{small-range}). The first observation is that the volume
of the percolating lumps strongly depends upon the cutoff imposed as is shown
on Fig.~\ref{fig:V-perc-small}. Moreover, the percolating lumps are present
only for $\Lambda_q \cdot a^4 \lesssim 7 \cdot 10^{-4}$ and disappear at larger values
of the cutoff. 
At the same time the power exponent $\alpha_V$ is also not universal as is clear from
Fig.~\ref{fig:alpha1} where we have plotted the distribution of lumps with small
volumes. We have also measured the function $\alpha_V(\Lambda_q)$ and it is presented
on Fig.~\ref{fig:alpha2}. The power exponent
$\alpha_V$ becomes cutoff independent only for $\Lambda_q \cdot a^4 > 10^{-3}$ thus
justifying the choice (\ref{Lambda-low}), (\ref{Lambda-high}) made earlier for the physical
range of the cutoff values.

To summarize, it seems crucial to ensure the universality of the lumps distribution
(and thus the very definition of the lumps) with respect to the {\it a priori} arbitrary
choice of the cutoff on the topological charge density. Once the universality is ensured
the lumps distribution follows sharply the cutoff independent power law (\ref{VQ-fit})
with rather remarkable value of the power exponents $\alpha_V \approx \alpha_Q \approx 3/2$. To the contrary
at small $\Lambda_q$ we found very peculiar lumps structure. Namely the number
of small lumps obeys the same power law (\ref{VQ-fit}) but
with much large power exponent which seems to be divergent in the limit $\Lambda_q \to 0$.
At the same time a few (typically two) percolating
lumps appear which are reminiscent to the global topological structures discovered
recently~\cite{Horvath-structures}.
However, it is unclear for us what is the physical significance
of the results obtained at very small $\Lambda_q$ since the lumps properties are
strongly cutoff dependent in this limit.

\subsubsection{Topological Density Correlation Function}
\label{qq-corr}

In this section we present our results for the topological charge density
correlation function $\langle q(0)q(x) \rangle \equiv \qq$ obtained on our $16^4$, $\beta=2.40$
configurations. However, let us clarify first our view on some theoretical points concerning
this correlator which were much elaborated in the recent studies~\cite{Horvath-qq}.

It has long been known (see Ref.~\cite{Seiler:2001je} and references therein)
that the correlation function $\qq$
can not be positive at non-zero physical distances. Confronted with the positivity
of the topological susceptibility $\chi = \langle Q^2 \rangle / V = \int \qq$
this implies a peculiar structure of $\qq$ correlation function,
which is to be negative for $x\ne 0$, while containing 
non-integrable singularity (contact term) at the origin such that
the susceptibility remains positive (in the context of $\mathrm{CP}^n$ models
this issue is also discussed in Ref.~\cite{Vicari:1999xx}).
In fact the negativity of $\qq$ correlator
motivated the recent discovery of the global topological structures~\cite{Horvath-structures}
since it implies that the topological charge could not be concentrated
in four-dimensional regions of finite physical size.
However, we do think that the contact term in $\qq$ correlator
is the clear sign of the perturbation theory to which the susceptibility
is believed to be unrelated.

To sort out the problem consider the instanton liquid model where the identification
of perturbative and non-perturbative contributions is
intuitively clean. At least at small non-zero distances we have $\qq_{pert}\sim - 1/|x|^8$,
while $\qq_{non-pert}$ is positive at small $x$ and rapidly drops down at distances comparable
with the characteristic instanton size (it could even become negative at larger $x$
if the instantons interaction is significant). If we suppose that the perturbative
and non-perturbative parts do not mix then the negativity requirement could be fulfilled
at least in principle.
The outcome is that the perturbation theory could overwhelm the non-perturbative part
and reproduce qualitatively the behavior of $\qq$ outlined above still giving vanishing
contribution to the susceptibility.
Then the actual question is how do we understand the topological charge density operator.
The negativity requirement could hold for perturbation theory sensitive (thus presumably local)
definitions of $q(x)$ and does not apply to non-perturbative ones.
Thus the problem we're discussing is essentially the same long standing problem of perturbative
and non-perturbative physics separation, which seems to have no unambiguous solution (see, e.g.,
Ref.~\cite{Zakharov:2003vd} for recent discussion).

A particular illustration of the above reasoning is provided by the overlap-based
construction of the topological charge~\cite{overlap}.
It is remarkable that once only a few lowest modes of the Dirac operator are considered
the topological charge density distribution has lumpy structure~\cite{DeGrand,Hip,Horvath:2002gk}
while the $\qq$ correlation function stays non-negative at all distances.
However, the inclusion of higher Dirac modes into the $q(x)$ definition
results in the appearance of some global ``sheets'' \cite{Horvath-structures}
percolating through all the lattice volume and carrying extremely large topological charge.
Simultaneously with increasing number
of modes the $\qq$ correlation function becomes negative and develops the positive
singularity at the origin~\cite{Horvath-qq}. The interpretation might be that the more modes are included
the more local the definition of topological density is and hence $q(x)$ becomes more sensitive
to the perturbation theory. Essentially for this reason we considered the effective topological
density (\ref{overlap-2}) in section~\ref{overlap-compare}, where the overlap-based definition
was compared with our construction.

It is remarkable that we also see qualitatively the same behavior with our
definition of the topological charge density. The border between
perturbative and non-perturbative regions is provided in our case by
the universality window with respect to $\Lambda_q$. We have seen that
the lumpy structure gradually changes with diminishing $\Lambda_q$
and becomes consisting of a few percolating lumps plus the divergent number of lumps having
UV-scale volume. However, we stress that not only
the window (\ref{Lambda-low}), (\ref{Lambda-high}) provides the separation of perturbative
and non-perturbative physics. Some degree of non-locality is built into our approach
from very beginning and in the spirit of the above discussion one could say that
our definition of the topological density is non-perturbative by construction
to which the negativity requirement does not apply. Thus we expect
that the $\qq$ correlation function in our case should behave roughly exponentially
\beq
\label{qq-fit}
\qq ~\sim~ \exp\{\, - |x|/\rho_\cL\,\}
\eeq
with correlation length $\rho_\cL$ corresponding to the characteristic lump size.

\begin{figure}[t]
\centerline{
\psfig{file=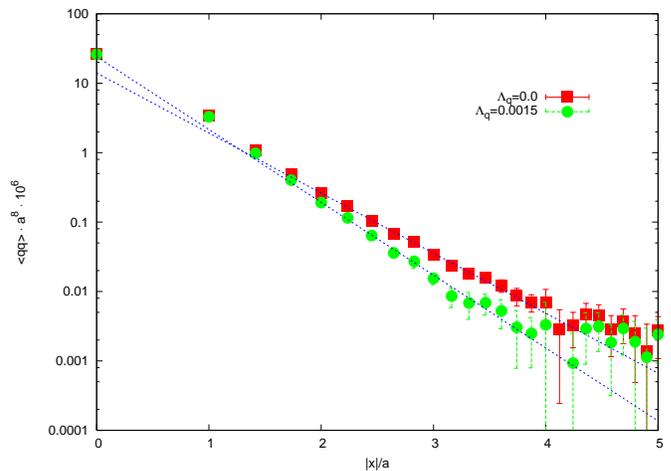,width=0.5\textwidth,silent=,angle=-90,clip=}
}
\caption{The topological charge density correlation function measured on $16^4$, $\beta=2.40$
lattices for $\Lambda_q \cdot a^4 = 0$ and $1.5 \cdot 10^{-3}$. Lines represent the best fits (\ref{qq-fit}).}
\label{fig:qq}
\end{figure}

The result of our measurements of the $\qq$ correlator on $16^4$, $\beta=2.40$
configurations is presented on Fig.~\ref{fig:qq}. It should be noted that within
the numerical errors $\qq$ is nowhere becoming negative and is compatible with
zero from $|x|=6 \cdot a$ onward. Therefore we plotted the $\qq$ correlator only in the range
$|x|/a \in [0;5]$ where data closely follows Eq.~(\ref{qq-fit}). In turn, the fitted
value of the characteristic lump size is given by 
\beq
\label{rho-lump-0}
\left. \rho_\cL\right|_{\Lambda_q = 0} ~=~ 0.50(2) \cdot a ~=~ 0.060(3) \mathrm{~fm}\,,
\eeq
where we have explicitly indicated that the data was taken without any cutoff imposed.
As far as the dependence upon the cuts (\ref{Lambda-low}), (\ref{Lambda-high}) and (\ref{volume-cut})
is concerned, it turns out that the $\qq$ correlator is independent on $V_{min}$ at non-zero separations
provided that $V_{min}$ is of order unity. Indeed, the dominant contribution to $\qq$ comes
from the extended regions of sign-coherent topological charge while the contribution
of small lumps $V(\cL_i) \approx 1$ averages to zero for $x\ne 0$. Note that the apparent
deviation of $\qq$ from (\ref{qq-fit}) at $|x|/a=0,1$ is due to the abundance of small lumps
and practically disappears once the cut $V_{min} \approx 1$ is applied.
The dependence upon $\Lambda_q$ could also be qualitatively established
since for increasing cutoff the size of the lump contributing significantly
to $\qq$ diminishes. In fact,  $\rho_\cL$ depends linearly on $\Lambda_q$
and for $\Lambda_q \cdot a^4 = 1.5 \cdot 10^{-3}$ it is given by
\beq
\label{rho-lump-1.5}
\left.\rho_\cL\right|_{\Lambda_q \cdot a^4 = 1.5 \cdot 10^{-3}} ~=~ 0.41(2) \cdot a ~=~  0.049(2) \mathrm{~fm}\,.
\eeq

To summarize, the topological charge correlation function $\qq$ respects the above investigated
lumpy structure of the topological charge bulk distribution. Namely, it falls exponentially with
distance, the corresponding characteristic size of the lumps is given by (\ref{rho-lump-0})
and decreases if the non-zero cutoff $\Lambda_q$ is applied, Eq.~(\ref{rho-lump-1.5}).
The quoted $\rho_\cL$ values are in qualitative agreement with lumps volume distribution 
we investigated previously, namely, it is rather small in lattice units respecting the ultraviolet
divergence present in (\ref{VQ-fit}), (\ref{VQ-fixed}).

\subsubsection{Scaling Check}
\label{scaling}

In this section we confront the data obtained at $\beta=2.40$ and $\beta=2.475$
in order to check the scaling properties of the quantities introduced earlier. 
First, consider the physical range of the cutoff $\Lambda_q$
which could be characterized by $\Lambda^*_q$, Eq.~(\ref{Lambda-ref}).
Performing the analogous calculations for $\beta=2.475$ we found that
\beq
\label{2.475-range}
\Lambda^*_q = (4 \div 8) \cdot 10^{-4} \cdot a^{-4} = (340(28)\mathrm{~MeV})^4\,,
\eeq
which is consistent with its value at $\beta=2.40$. Thus the cutoff introduced
to define the lumps appears to be physical quantity with continuum value around
$(300\div 350\mathrm{~MeV})^4$. Note that the range of $\Lambda_q$ where $\diff_{\Lambda_q} D_\cL$
approximately vanishes is notably wider at $\beta=2.475$ then is was at $\beta=2.40$.

As far as the power exponents $\alpha_V$ and $\alpha_Q$  are concerned,
at $\beta=2.475$ they turn out to be also compatible with each other. In particular,
$\alpha_V$ values at the boundaries of the range (\ref{2.475-range}) are given by
\beqn
\left.\alpha_V\right|_{\Lambda_q \cdot a^4 = 4 \cdot 10^{-4}} & = & 1.62(14)\,, \\
\left.\alpha_V\right|_{\Lambda_q \cdot a^4 = 8 \cdot 10^{-4}} & = & 1.56(12) \nonumber
\eeqn
and are consistent with the conjecture (\ref{VQ-fixed}). On the other hand,
the topological susceptibility is lumps saturated for $\Lambda_q \cdot a^4 < 10^{-3}$
which is even a bit beyond the upper bound quoted in (\ref{2.475-range}).
Therefore the estimate of the universality window (\ref{2.475-range}) is in fact rather
conservative.

Finally, we considered the topological density correlation function at $\beta=2.475$.
It turns out that it is again non-negative within the numerical errors and falls off
exponentially with distance. The corresponding characteristic lump size at $\beta=2.475$ 
and $\Lambda_q = 0$ is
\beq
\label{rho-lump1}
\left.\rho_\cL\right|_{\Lambda_q = 0} ~=~ 0.57(2) \cdot a ~=~ 0.052(2) \mathrm{~fm}\,,
\eeq
which disagrees with (\ref{rho-lump-0}) and  shows instead that $\rho_\cL (\Lambda_q = 0)$
is almost constant in lattice units. However, this should not be surprising since $\Lambda_q = 0$
is definitely outside the physically acceptable region. On the other hand, the characteristic lump size
measured at $\Lambda_q \cdot a^4 = 6 \cdot 10^{-4}$ turns out to be
\beq
\label{rho-lump2}
\left.\rho_\cL\right|_{\Lambda_q \cdot a^4 = 6 \cdot 10^{-4}} ~=~ 0.50(1) \cdot a ~=~ 0.046(1) \mathrm{~fm}
\eeq
and is compatible with (\ref{rho-lump-1.5}).
At present it is too early to speculate about the observed scaling in the topological
charge density correlation function, obviously more data is needed to quantify the issue.

\section{Conclusions}
\label{conclude}
In this paper we considered the alternative definition of the topological charge
density in pure SU(2) lattice gauge model. Generically the idea is to exploit
the well known close connection between $\HP{n}$ $\sigma$-models and SU(2) Yang-Mills
theory, which was proved to be very useful in the investigations of Yang-Mills fields
topology. The usual approach in the past was to consider the $\sigma$-model instantons
and to induce the corresponding SU(2) gauge potentials which realize the topologically non-trivial
configurations in the gauge theory. We tried instead to read the above relation
in the opposite way, namely, to find the unique $\HP{n}$ $\sigma$-model fields which
are closest to the given SU(2) gauge background.
In the continuum limit and for the smooth fields we are confident that our construction
is identical to any other way of the topological charge calculation. However, considered
within the regularized quantum field theory the $\HP{n}$ $\sigma$-model approach
obtains a separate significance and provides an alternative definition
of the topological charge on the lattice. Moreover, the $\sigma$-model construction
naturally leads to the unambiguous definition of the topological charge density
with clean geometrical meaning.

The actual realization of the above program brought out a wealth of technical issues
which, we believe, were adequately addressed in the paper. Moreover, the algorithm was tested in various
circumstances and, in particular, it was compared with overlap-based topological density
definition. In the latter case we found a strong evidences that both approaches are consistent
with each other.

As far as the topology of SU(2) gauge fields are concerned the results we obtained are
at least in qualitative agreement with the literature. In particular, the topological
susceptibility measured by our method seems to be perfectly scaling and coincides with
its conventional value.  Note that at present there is no commonly accepted picture of
the topological charge bulk distribution. Here our approach provides the unique
investigation tool and our main results could be summarized as follows.

We confirm the lumpy structure of the topological charge bulk distribution
discovered in the previous studies and assert that these lumps are not compatible
with instantons. Moreover, for carefully chosen parameters entering the lumps definition
we found that their volume distribution seems to obey rather remarkable power low
$N(V_\cL) \sim V_\cL^{-3/2}$, while their topological charge is likely to be proportional
to the lumps volume with distribution being essentially the same.
At the same time the lumps do saturate the topological
susceptibility and in the first approximation the topological charge is indeed
concentrated in the lumps only.

Note that the lumps volume distribution naively implies that the characteristic size
of the lumps is given by the UV scale. On the other hand, it could be probed by
the topological density correlation function $\qq$ provided that we could separate
the perturbative contribution from it. We argued that our topological density construction
is inherently non-perturbative and its correlation function generically 
does not obey the negativity requirement.
In turn the data indicate strongly that the $\qq$ correlator falls off exponentially,
the correlation length being indeed of order half the lattice spacing. However, it increases
with diminishing spacing so that the possible scaling is not excluded.
In either case the characteristic lumps size extracted from $\qq$ correlation function
is definitely smaller then $0.1\mathrm{~fm} \approx 2\mathrm{~GeV}^{-1}$ and is worth to
be compared with the approximate location of $\Lambda_q$ universality window $(300\div 350\mathrm{~MeV})^4$.
The scale of few hundred MeV is inherent the gauge fields topology, for instance,
it is built into phenomenological instantonic models.
In our approach this scale is the lower bound on the characteristic topological density
in the lumps and separates the lumps from the UV noise present in $q(x)$.
Thus at zero approximation the lumps could be viewed as highly localized
$\rho_\cL \lesssim 2\mathrm{~GeV}^{-1}$ and 'hot' $q_\cL \gtrsim [300 \mathrm{~MeV}]^4$ bumps
in the topological density, where $\rho_\cL$ and $q_\cL$ are the characteristic lumps size and density
correspondingly. Therefore they could strongly
influence the microscopic structure of low-lying Dirac eigenmodes.
In particular, the lumps smallness could explain the unusual localization properties
of low Dirac eigenmodes discovered recently~\cite{Gattringer,localization,Gubarev:2005jm},
while their characteristic topological density could reveal itself in the Dirac modes mobility
edge~\cite{Gubarev:2005jm}.

\section*{Acknowledgments}
\noindent
The authors are grateful to prof. V.I.~Zakharov and to the members
of ITEP lattice group for stimulating discussions. The work was partially
supported  by grants RFBR-05-02-16306a, RFBR-05-02-17642, RFBR-0402-16079 and
RFBR-03-02-16941. F.V.G. was partially supported by INTAS YS grant 04-83-3943.


\end{document}